\documentclass[twocolumn,prl,superscriptaddress,showpacs,preprintnumbers,amsmath,amssymb]{revtex4}

\usepackage{graphicx}% Include figure files
\usepackage{bm}% bold math

\expandafter\ifx\csname package@font\endcsname\relax\else
 \expandafter\expandafter
 \expandafter\usepackage
 \expandafter\expandafter
 \expandafter{\csname package@font\endcsname}%
\fi

\newcommand{\fp}[1]{(#1)} %figure part
\newcommand{\fpc}[1]{(#1)} %figure part in caption
\newcommand{\fig}[1]{Fig.\ \ref{#1}}

\newcommand{\tmax}{\theta_\mathrm{max}}
\newcommand{\tr}{\theta_r}
\newcommand{\Dt}{\Delta\theta}
\newcommand{\DE}{\Delta E}
\newcommand{\oil}[1]{$\tau = #1$\,\%}
\newcommand{\bead}[1]{$d = #1$\,mm}
\newcommand{\rot}[1]{$\Omega = #1$\,rpm}
\newcommand{\avg}[1]{\langle #1 \rangle}
\newcommand{\stddev}[1]{\sigma(#1)}
\newcommand{\huv}{h_{u,v}}

\newcommand{\redfig}{{\em This is a reduced resolution figure }\protect\cite{fullfigs}. }

\begin{document}

\title{Avalanche Dynamics in Wet Granular Materials}

\author{P. Tegzes}
\affiliation{Department of Physics and Materials Research Institute, %
Pennsylvania State University, University Park PA 16802}
\affiliation{Department of Biological Physics, E\"otv\"os Lor\'and University, %
1A P\'azm\'any s\'et\'any, Budapest, Hungary 1117}
\author{T. Vicsek}
\affiliation{Department of Biological Physics, E\"otv\"os Lor\'and University, %
1A P\'azm\'any s\'et\'any, Budapest, Hungary 1117}
\author{P. Schiffer}
\affiliation{Department of Physics and Materials Research Institute, %
Pennsylvania State University, University Park PA 16802}

\date{\today}

\begin{abstract}
A detailed characterization of avalanche dynamics of wet granular media in a
rotating drum apparatus is presented. The results confirm the existence of the
three wetness regimes observed previously: the granular, the correlated and the
viscoplastic regime. These regimes show qualitatively different dynamic
behaviors which are reflected in all the investigated quantities. We discuss
the effect of interstitial liquid on the characteristic angles of the material
and on the avalanche size distribution. These data also reveal logarithmic
aging and allow us to map out the phase diagram of the dynamical behavior as a
function of liquid content and flow rate. Via quantitative measurements of the
flow velocity and the granular flux during avalanches, we characterize novel
avalanche types unique to wet media. We also explore the details of
viscoplastic flow (observed at the highest liquid contents) in which there are
lasting contacts during flow, leading to coherence across the entire sample.
This coherence leads to a velocity independent flow depth at high rotation
rates and novel robust pattern formation in the granular surface.
\end{abstract}

\pacs{45.70.-n, 45.70.Ht, 45.70.Mg}

\maketitle

\section{Introduction}

While most research on the physics of granular media has focused on dry
grains, the presence of even microscopic quantities of interstitial liquid can
have profound effects on the physical behavior
\cite{Nedderman92,Albert97,Hornbaker97,Halsey98,Mason99,Bocquet98,%
Fraysse97,Samadani00,Samadani01}. Capillary forces lead to
cohesion which greatly enhances the stability of wet samples, and
several recent studies have investigated liquid induced effects on
the static properties
\cite{Albert97,Hornbaker97,Halsey98,Mason99,Bocquet98,Fraysse97}.
In previous studies of the repose angle of wet granular media
using a draining crater apparatus \cite{Tegzes99}, we identified
three fundamental regimes as a function of the liquid content. The
{\em granular regime} at very low liquid contents is dominated by
the motion of individual grains; in the {\em correlated regime}
corresponding to intermediate liquid contents, a rough surface is
formed by the flow of separated clumps; and the repose angle of
very wet samples results from smooth, cohesive flow with {\em
viscoplastic} properties. The addition of liquid also
qualitatively changes the dynamic behavior of granular media, as
was observed qualitatively in our angle of repose measurements and
also revealed in recent segregation studies
\cite{Samadani00,Samadani01}.  A quantitative characterization of
the {\em dynamics} of wet granular flow, however, has thus far
been missing. In this paper we study the avalanche dynamics and
flow properties of wet granular materials to investigate how the
gradual addition of a small amount of liquid influences the
dynamic properties of granular surface flow, and we find that the
three fundamental regimes observed earlier are also reflected in
the present experiment. Some of the results presented here have
been published elsewhere \cite{Tegzes02}.

Granular surface flow occurs when the inclined surface of a granular medium
loses its stability against gravitational force. There are several experimental
methods to investigate such flow, e.g. by
gradually tilting a granular sample, which leads to an avalanche when the surface
angle exceeds a critical value $\tmax$ \cite{Rajchenbach02}. Alternatively, if
the surface angle is close to $\tmax$, avalanches can be triggered by 
point-like perturbations \cite{Daerr99}. The inclination of the surface can also
be increased by the addition of grains to the top of the pile
\cite{Held90,Lemieux00} or by slowly decreasing the size of the supporting base
\cite{Jia00}.

In our measurements we employ a rotating drum apparatus (a cylindrical chamber
partly filled with a granular medium and rotated around a horizontal axis)
\cite{Rajchenbach90,Liu91}, which tilts the sample in a highly controlled and
reproducible manner and offers several benefits. The most important advantage
of a rotating drum is that at low rotation rates it allows for the observation of many avalanches
without the need to change the sample. After each avalanche the medium remains
at rest relative to the drum while its surface angle is slowly increased by
rotation, until it reaches $\tmax$ again. Then another avalanche occurs, and
the process starts over. Thus it is possible to record hundreds of avalanches,
which is essential for performing a statistical analysis or for averaging out
noise-like fluctuations in dynamical data. In addition to the avalanches
observed at low rotations rates, the rotating drum also allows the
investigation of continuous flow down the slope at higher rotation rates.

The sections of this paper cover different aspects of our rotating drum
experiments. In Section \ref{sec_exp} we describe the experimental setup.
Section \ref{sec_angle} addresses some issues explored in previous studies of
cohesive granular media in a rotating drum
\cite{Bocquet98,Fraysse99,Nase01,Viturro01}, focussing on the surface angles
of the medium before and after avalanches, and describing a statistical analysis
of avalanche size based on these angles. In Section \ref{sec_phdiag} we
investigate the transition between avalanches and continuous flow as a function
of the liquid content, and map out the phase diagram of the system. Then a
short section (\ref{sec_morph}) is devoted to the morphology of the surface. In
the rest of the paper we focus on characterizing the {\em dynamics of cohesive
flow}. In Section \ref{sec_avadyn} we quantitatively investigate the flow
dynamics during avalanches at different liquid contents by analyzing the time
evolution of the averaged surface profile obtained from hundreds of avalanche
events. In Section \ref{sec_flowdyn} we analyze the dynamical properties of the
continuous flow phase occurring at faster rotation. We pay special attention to
the nature of the {\em viscoplastic} flow, observed at the highest liquid
contents which displays unique characteristics associated with coherent motion
over the entire granular surface. In the final section we summarize and discuss
our results.

%%%%%%%%%%%%%%%%%%%%%%%%%%%%%%%%%%%%%%%%%%%%%%%%%%%%%%%%%%%%%%%%%%%%%%
%%%%                                                              %%%%
\begin{figure}%
 \includegraphics[clip,width=\columnwidth]{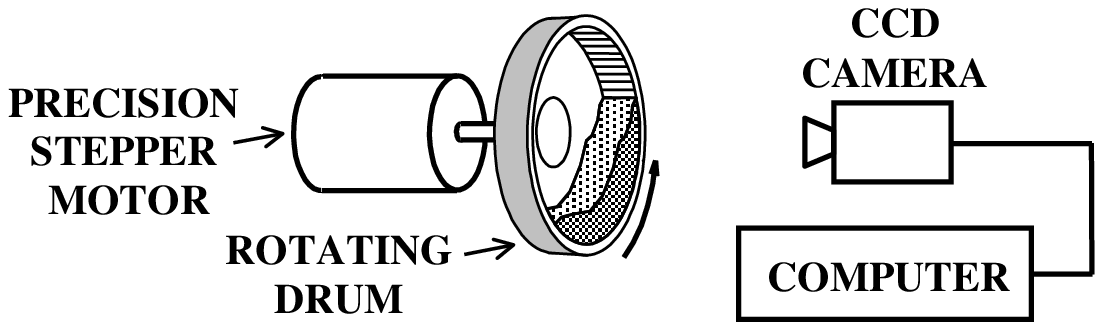}%
 \caption{Sketch of the experimental setup. }%
 \label{fig_exp_sketch}%
\end{figure}%
%%%%                                                              %%%%
%%%%%%%%%%%%%%%%%%%%%%%%%%%%%%%%%%%%%%%%%%%%%%%%%%%%%%%%%%%%%%%%%%%%%%

\section{Description of the apparatus }
\label{sec_exp}

We studied glass spheres thoroughly mixed with small quantities of
hydrocarbon oil. The viscosity of the oil used was $0.27$ poise,
and its surface tension was $0.02$\,N/m. The
liquid content varied between \oil{0.001} and $5\,\%$ of the void
volume. In this regime the flow of oil due to gravity can be
neglected. Measurements were performed on three sizes of beads,
with diameters $d=0.9\,$mm$\,\pm\,11\,\%$,
$d=0.5\,$mm$\,\pm\,20\,\%$ and $d=0.35\,$mm$\,\pm\,15\,\%$.  Note the the beads differed not only in their size, but also presumably in their microscopic surface structures, which are difficult to characterize or control. Nonetheless, most of the qualitative behavior is reproduced in the different grain samples,  those differences which are significant are mentioned below.  

The rotating drum (\fig{fig_exp_sketch}) was made of thick
plexiglas, but the vertical walls were lined with glass plates in
order to minimize electrostatic effects. To prevent slips along
the circumference of our drum, we inserted a hollow, thin aluminum
cylinder with a rough inner surface into the drum. The inner
diameter of the drum was $16.8$ cm, its width was $3.2$ cm, and
the granular filling was $30\,\%$. By performing measurements in a
thinner ($2$ cm) drum we verified that, while wall effects are not
negligible, they do not appear to modify the qualitative behavior.

The drum was rotated by a computer-controlled precision stepping motor which
provided an extremely stable rotation rate that could be varied in the range of
$\approx 0.003-30$ rotation per minute (rpm). The step size of rotation was
$0.025^\circ$. The vibrations originating from the stepper motor were damped by
multiple lead bricks attached to the apparatus.

%%%%%%%%%%%%%%%%%%%%%%%%%%%%%%%%%%%%%%%%%%%%%%%%%%%%%%%%%%%%%%%%%%%%%%
%%%%                                                              %%%%
\begin{figure}
 \includegraphics[clip,width=\columnwidth]{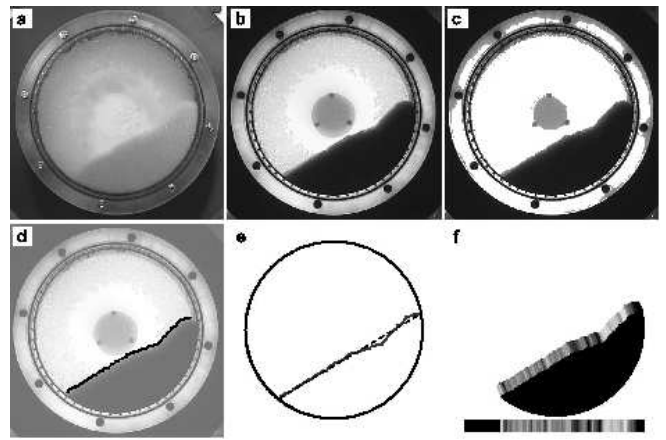}
\caption{\redfig Steps of image processing. \fpc{a} Snapshot taken without
background illumination. The wet grains sticking to the glass
walls make the surface profile hardly visible. \fpc{b} The same
picture but with a circular lamp turned on in the background. In
diffuse background illumination the monolayer of beads on the wall
is much brighter than the $3.2$ cm thick sample. \fpc{c} A
brightness threshold is applied to the image to make the beads on
the walls disappear. \fpc{d} Test of the algorithm: the surface
data (black line) superimposed over a greyed version of the
original photograph. \fpc{e} The surface profile with a fitted
straight line determining the overall surface angle ($\theta$).
\fpc{f} Pseudo-3dimensional representation of the surface profile,
shading indicates local angle ($\alpha(x)$). This shading may be
used to represent a given surface profile in a single line (see
the horizontal stripe at the bottom).}
 \label{fig_drum_proc}
\end{figure}
%%%%                                                              %%%%
%%%%%%%%%%%%%%%%%%%%%%%%%%%%%%%%%%%%%%%%%%%%%%%%%%%%%%%%%%%%%%%%%%%%%%

The experiment was recorded by a charge coupled device video camera interfaced to a
computer which could analyze the spatiotemporal evolution of the
surface profile (height variations in the axial direction were
negligible). The steps of image processing are demonstrated in
\fig{fig_drum_proc}\fp{a-c}. When the grains are wet, the beads
sticking to the glass walls make the surface hardly visible
(\fig{fig_drum_proc}\fp{a}). We overcame this problem by using
background illumination (\fig{fig_drum_proc}\fp{b}) and applying a
brightness threshold (\fig{fig_drum_proc}\fp{c}) on the image.
We found the surface profile $h(x)$ by an algorithm based on
the detection of points of maximum contrast \cite{Tegzes02a}. The
resulting surface data are presented in \fig{fig_drum_proc}\fp{d-f}
in different ways. \fig{fig_drum_proc}\fp{d} shows a test of the
algorithm: the detected surface is superimposed on the photograph
of the system as a thin black line. The limitations of our camera
and computer gave us a resolution of $\delta t = 0.03$ seconds and
$\delta x = 0.5$\,mm. \fig{fig_drum_proc}\fp{e} presents the
$h(x)$ curve together with a straight line that was fitted to the
curve in a rotation-invariant way: the squared sum of the {\em
geometrical distances} of the surface point from the line was
minimized. The inclination angle of this line gives the {\em
overall angle} of the surface. This is the generalized version of
the surface angle that has typically been used to describe the
flat surface of dry samples. Finally, \fig{fig_drum_proc}\fp{f}
shows the surface profile with a brightness-coded representation
of the local angle ($\alpha(x)=\arctan \left[{d \over dx}
h(x)\right]$) in the third dimension. The stripe at the bottom is
a concise way of displaying this surface profile using the same
shading. In sections \ref{sec_avadyn} and  \ref{sec_flowdyn} we
use this representation of the surface topography to visualize the
time evolution of the surface by placing the stripes corresponding
to successive surface profiles under each other.

\section{Dynamic properties from surface angle measurements}
\label{sec_angle}

The phenomenology of wet granular materials is much richer than
that of dry samples. The attractive forces between the particles
lead to correlated motion of the grains and qualitatively new
avalanche dynamics. These new types of avalanches form rough and
highly variable surfaces. In order to capture the most fundamental
features of our system, however, we will initially neglect these
surface features and examine only the {\em overall surface angle}
(\fig{fig_drum_proc}\fp{e}) to describe the observed surfaces with
a single number. This approach allows us to compare our results
directly to the behavior of dry materials, where the surface is
virtually flat and is completely described by its inclination
angle.

%%%%%%%%%%%%%%%%%%%%%%%%%%%%%%%%%%%%%%%%%%%%%%%%%%%%%%%%%%%%%%%%%%%%%%
%%%%                                                              %%%%
\begin{figure}
 \includegraphics[clip,width=\columnwidth]{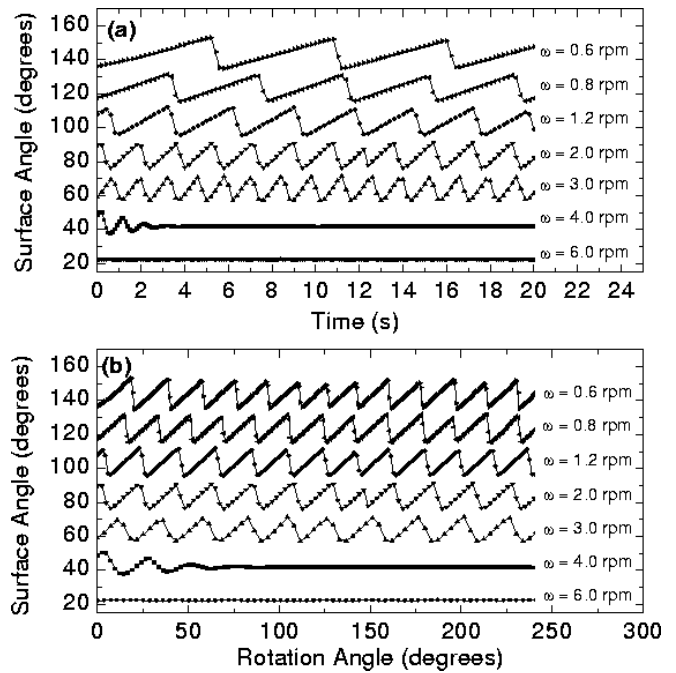}
\caption{\redfig The temporal variation of the surface angle at various
rotation rates \fpc{a} as a function of time, \fpc{b} as a
function of the rotation angle of the drum. These data were taken
using the large \bead{0.9} beads, the liquid content was
\oil{0.009}. Note that the curves are shifted by increments of $20^\circ$ for clarity. }
 \label{fig_ang_raw}
\end{figure}
%%%%                                                              %%%%
%%%%%%%%%%%%%%%%%%%%%%%%%%%%%%%%%%%%%%%%%%%%%%%%%%%%%%%%%%%%%%%%%%%%%%

%%%%%%%%%%%%%%%%%%%%%%%%%%%%%%%%%%%%%%%%%%%%%%%%%%%%%%%%%%%%%%%%%%%%%%
%%%%                                                              %%%%
\begin{figure}
 \includegraphics[clip,width=7cm]{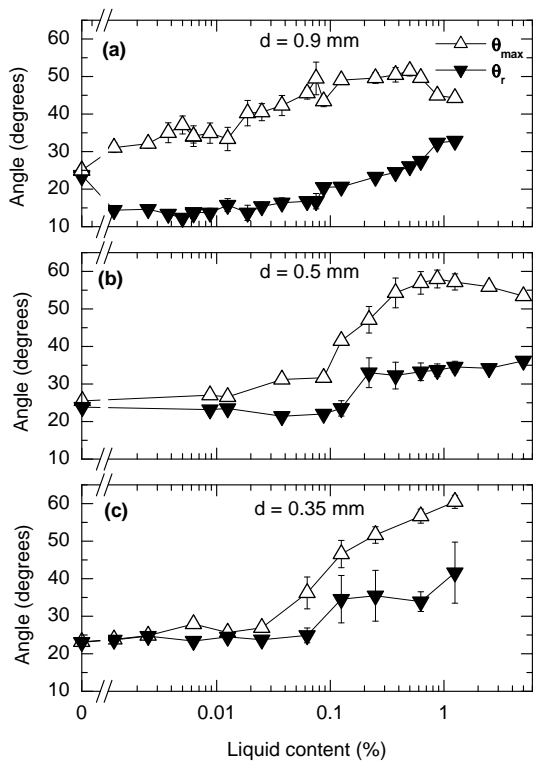}
\caption{The dependence of the maximum angle $\tmax$ and the
repose angle $\tr$ on the liquid content for 3 different bead
sizes. \fpc{a} \bead{0.9}, \rot{0.12}, \fpc{b} \bead{0.5},
\rot{0.12}, \fpc{c} \bead{0.35}, \rot{0.03} (a lower rotation rate
was chosen, because at \rot{0.12} the smallest beads exhibit
continuous flow for the lowest liquid contents). }
 \label{fig_maxmin}
\end{figure}
%%%%                                                              %%%%
%%%%%%%%%%%%%%%%%%%%%%%%%%%%%%%%%%%%%%%%%%%%%%%%%%%%%%%%%%%%%%%%%%%%%%

\fig{fig_ang_raw} shows the time evolution of the surface angle
for various rotation rates. Similarly to the dry materials we
observe discrete avalanches at low rotation rates. We find that
$\theta(t)$ in this regime is a typical sawtooth-shaped function
looking very similar to the force-displacement curve of stick-slip
processes, which is not very surprising since stick-slip-like
behavior is quite frequent in granular materials
\cite{Nasuno97,Albert00,Albert01,Cain01}. The local maximum and
minimum points of $\theta(t)$ correspond to the maximum angle of stability 
($\tmax$) and the repose angle ($\tr$)
respectively. Note, that if we plot the surface angle as a
function of the rotation angle of the drum, then the increasing
segments of the curve are straight lines with a slope of unity. If
we gradually increase the angular velocity of the drum, then at a
given rotation rate (\rot{4} in \fig{fig_ang_raw}) the flow
becomes continuous, and after some transients the surface angle
becomes constant. In this section we consider the avalanching
regime (at \rot{0.12}, except where noted otherwise) and
investigate how the characteristic angles and the avalanche size
distribution depends on the liquid content and the rotation rate.
The transition to continuous flows is described in the subsequent
section.

\subsection{The maximum and the repose angle}
\label{sec_maxmin}

\fig{fig_maxmin} shows the maximum angle just before an avalanche, 
$\avg{\tmax}$, and the angle of repose after an avalanche, $\avg{\tr}$, 
as a function of liquid content
for the three different bead sizes (averaged
over several hundred avalanches). The measured angles have
distributions of finite width, the error bars in the figure
indicate their standard deviation. These characteristic angles
reveal essential properties of the material: $\tmax$ reflects the
stability of the sample while $\tr$ is also related to the
dissipation during the avalanche events. The various features of
these curves reflect changes in the dynamical behavior
corresponding to the three regimes mentioned in the introduction,
which are described in this section through qualitative
observations, and are more quantitatively characterized in Section
\ref{sec_avadyn}.

The curves corresponding to the smaller bead sizes \bead{0.5}
(\fig{fig_maxmin}\fp{b}) and \bead{0.35} (\fig{fig_maxmin}\fp{c})
are quite similar. At the lowest liquid contents, very small
avalanches are observed. This is the granular regime, where the
behavior is qualitatively similar to that of dry granular
materials. Due to the interparticle cohesive forces, $\tmax$ in
this regime is larger than the value measured in dry samples. On
the other hand, $\tr$ hardly changes, presumably because the
presence of interstitial liquid does not significantly alter the
dissipation in the fluidized layer of rolling beads.

At higher liquid contents, in the correlated regime, the increased
cohesive force results in a larger increase in $\tmax$. In this
regime the beads on the top surface are more effectively
stabilized by the cohesive forces, thus failure leading to an
avalanche occurs in the bulk of the sample, and the avalanche flow
includes the motion of correlated clumps. We find that $\tr$ also
displays a sudden increase, but at somewhat higher liquid contents
than the increase in $\tmax$.  This offset is possibly due to the
motion of the clump fluidizing the underlying layer of beads, thus individual rolling grains help determine the final slope as
in the granular regime. At higher liquid contents, the cohesive
and viscous forces would prevent such fluidization leading to the
increase in $\tr$.

At the highest liquid contents, $\tmax$ slightly decreases in
\fig{fig_maxmin}\fp{a} and \fp{b}. This decrease in the stability
of the material, also observed in our previous studies of the
repose angle using a different technique \cite{Tegzes99}, is
probably related to lubrication effects decreasing the static
friction among the grains. In this regime the nature of the
avalanches changes again: the material is transported by coherent
flow extending over the whole surface of the sample. We did not
observe this regime for the \bead{0.35} beads, possibly because, for those grains, the cohesive forces are much stronger  relative to
their weight.  Careful examination of \fig{fig_maxmin} reveals a
few other differences between the different grains studied such as
the relatively large avalanches exhibited by the \bead{0.9} beads
even at very low liquid contents.  These differences are difficult
to interpret, since we only examined three grain samples which
varied not only in their size, but presumably also in their
microscopic surface properties.

\subsection{Statistics of avalanche sizes}

Thus far we have restricted our discussion to the average values
$\avg{\tmax}$ and $\avg{\tr}$. However, the characteristic angles
are influenced by the microstructure of the sample, i.e., the
actual configuration of the grains, which leads to inherent
fluctuations and a non-trivial distribution of the measured
quantities. The analysis of these distributions may provide
insight into the internal dynamics of the material. In particular
the statistics of avalanche sizes has attracted much recent
attention due to its possible connection to self organized
criticality
\cite{Bak87,Rajchenbach90,Held90,Liu91,Bretz92,Frette96,Jia00,Viturro01}.

%%%%%%%%%%%%%%%%%%%%%%%%%%%%%%%%%%%%%%%%%%%%%%%%%%%%%%%%%%%%%%%%%%%%%%
%%%%                                                              %%%%
\begin{figure}
 \includegraphics[clip,width=7cm]{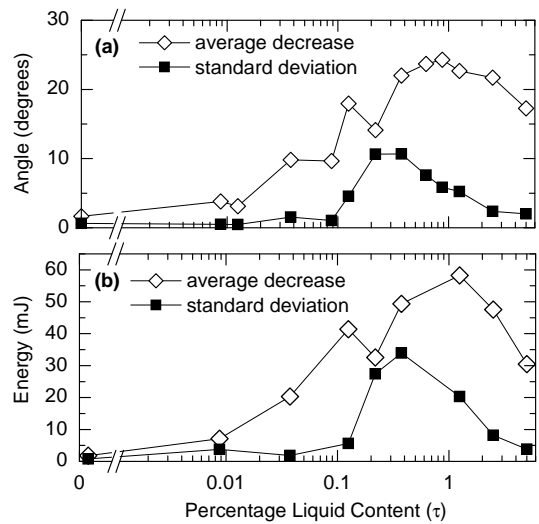}
 \caption{The average avalanche size and the width of the avalanche size
distribution as a function of liquid content, calculated with two
different methods. \fpc{a} Decrease in the surface angle: $\avg{\Dt}$,
$2\stddev{\Dt}$, \fpc{b} Decrease in the potential energy: $\avg{\DE}$,
$2\stddev{\DE }$. The data were taken using medium size beads,
\bead{0.5}, \rot{0.12}.}
 \label{fig_avastat_erg}
\end{figure}
%%%%                                                              %%%%
%%%%%%%%%%%%%%%%%%%%%%%%%%%%%%%%%%%%%%%%%%%%%%%%%%%%%%%%%%%%%%%%%%%%%%

%%%%%%%%%%%%%%%%%%%%%%%%%%%%%%%%%%%%%%%%%%%%%%%%%%%%%%%%%%%%%%%%%%%%%%
%%%%                                                              %%%%
\begin{figure}
 \includegraphics[clip,width=\columnwidth]{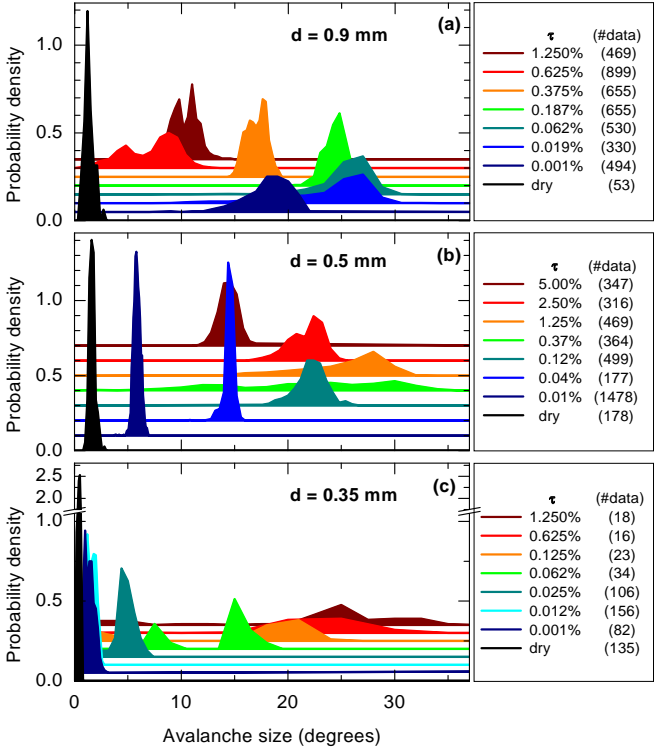}
\caption{(color) Distribution of the avalanche size ($\Dt$) for various
liquid contents.  \fpc{a} Large beads, \bead{0.9}, \rot{0.12},
\fpc{b} medium beads, \bead{0.5}, \rot{0.12}, \fpc{c} small beads,
\bead{0.35}, \rot{0.03}. The righthand column shows the number of avalanches
included for each sample - note that the statistics are much better for the
large and medium beads than for the smaller beads.}
 \label{fig_avadistr}
\end{figure}
%%%%                                                              %%%%
%%%%%%%%%%%%%%%%%%%%%%%%%%%%%%%%%%%%%%%%%%%%%%%%%%%%%%%%%%%%%%%%%%%%%%

%%%%%%%%%%%%%%%%%%%%%%%%%%%%%%%%%%%%%%%%%%%%%%%%%%%%%%%%%%%%%%%%%%%%%%
%%%%                                                              %%%%
\begin{figure}
 \includegraphics[clip,width=\columnwidth]{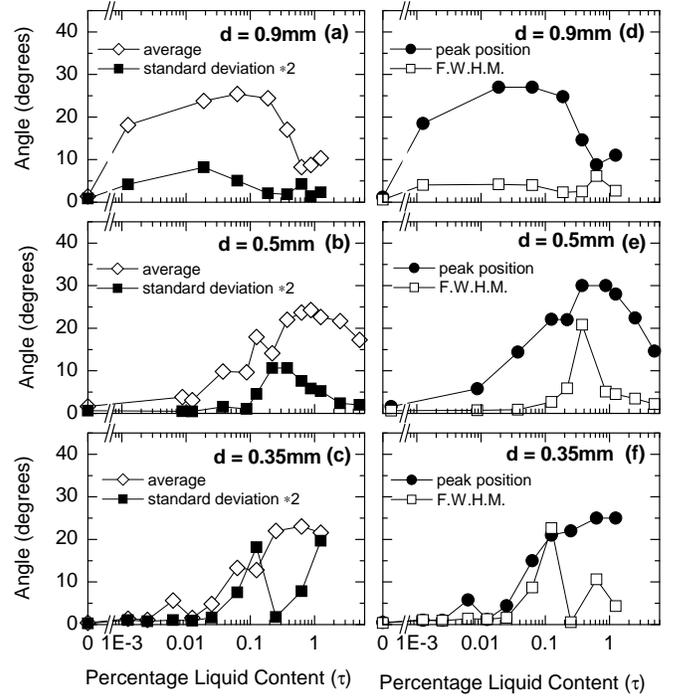}
\caption{Statistical measures of the avalanche size distributions
presented in \protect\fig{fig_avadistr}, for the 3 different bead
sizes as indicated in the figure. \fpc{a-c} The average angle
drop: $\avg{\Dt}$, and the standard deviation $2\stddev{\Dt}$ (see
text). \fpc{d-f} The most probable avalanche size, {\em i.e.} the
maximum of $P(\Dt)$ ("peak position"), and the width of the peaks
(full width at half maximum, "F.W.H.M."). }
 \label{fig_avastat}
\end{figure}
%%%%                                                              %%%%
%%%%%%%%%%%%%%%%%%%%%%%%%%%%%%%%%%%%%%%%%%%%%%%%%%%%%%%%%%%%%%%%%%%%%%

%%%%%%%%%%%%%%%%%%%%%%%%%%%%%%%%%%%%%%%%%%%%%%%%%%%%%%%%%%%%%%%%%%%%%%
%%%%                                                              %%%%
\begin{figure}
 \includegraphics[clip,width=\columnwidth]{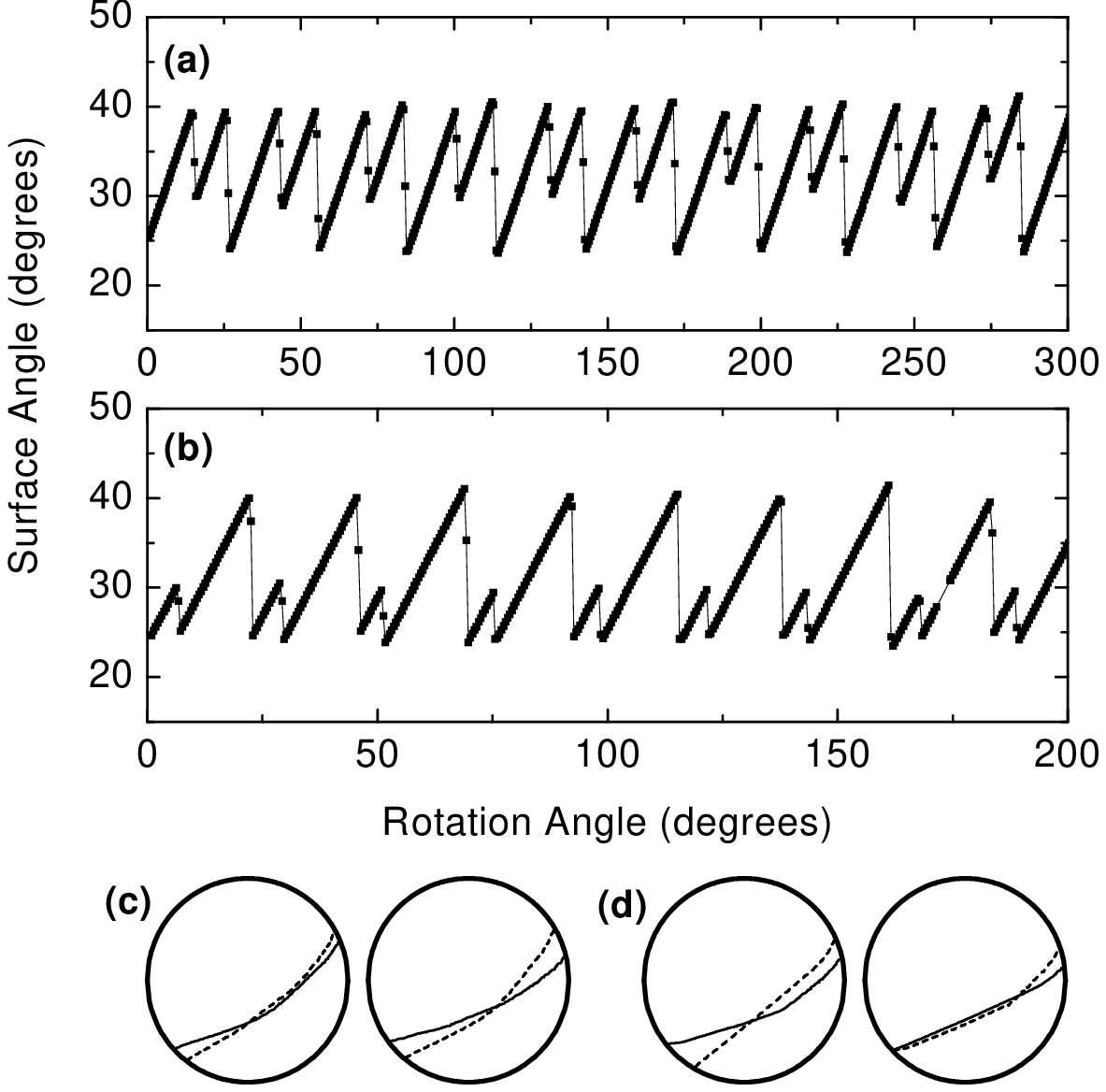}
\caption{Alternating small and large avalanches. \fpc{a}
\bead{0.5}, \oil{0.125}, \rot{0.4}, \fpc{b} \bead{0.35},
\oil{0.062}, \rot{0.15}. \fpc{c} and \fpc{d} The initial (dashed
line) and final (solid line) surface profiles for two successive
avalanches. The parameters are the same as in \fpc{a} and \fpc{b}
respectively.  Unfortunately, due to the fluctuations in the system these patterns of behavior were difficult to create reproducibly. }
 \label{fig_alternate}
\end{figure}
%%%%                                                              %%%%
%%%%%%%%%%%%%%%%%%%%%%%%%%%%%%%%%%%%%%%%%%%%%%%%%%%%%%%%%%%%%%%%%%%%%%

In this section, we analyze the statistical distribution of
avalanche sizes. The simplest quantity that describes the size of
an avalanche is $\Dt = \tmax - \tr$, which has been widely used to
characterize avalanches among dry grains in a rotating drum
\cite{Rajchenbach90,Evesque91}. In \fig{fig_avastat_erg}\fp{a} we
plot the average avalanche size
\begin{equation}
\avg{\Dt} = \frac{1}{N}\sum_{i=1}^N\Dt_i
\end{equation}
(where $\Dt_i$, $i=1..N$ is the angle drop in the $i^{th}$ avalanche) and the
width of the avalanche size distribution
\begin{equation}
\stddev{\Dt}=\sqrt{\frac{N\sum_{i=1}^N\Dt_i^2-(\sum_{i=1}^N\Dt)^2}{N(N-1)}}
\end{equation}
for the \bead{0.5} beads. (Note that we do not assume that the
distribution is Gaussian, we just use the standard deviation as a
measure of the width of the distribution.) Since the definition of
the surface angle is somewhat arbitrary in the case of our
surfaces which can be both rough and have global curvature (as
discussed below), we also define an alternative measure of the
avalanche size, $\DE$, which equals the difference in the
potential energy of the sample just before and just after the
avalanche \cite{Baumann96}. The analogous quantities $\avg{\DE}$
and $2\stddev{\DE}$ are plotted in \fig{fig_avastat_erg}\fp{b}.
The graphs indicate that the two different measures of the
avalanche size yield almost equivalent results, thus we use the
standard $\Dt$ as the measure of the avalanche size in the
following discussion.  In order to characterize the distribution
in more detail, in \fig{fig_avadistr} we plot the measured
probability densities $P(\Dt)$ for several liquid contents for the
3 different bead sizes, where $P(\Dt)d\theta$ is the probability
for the avalanche size to fall between $\Dt$ and $\Dt+d\theta$ (we
estimate $P(\Dt)$ from histograms). In \fig{fig_avastat} we also
plot several statistical measures characterizing these
distributions including the peak position (the maximum of
$P(\Dt)$) and the FWHM (full width of the peak at half of the
maximum value).

The statistical properties of the avalanches shown in
\fig{fig_avadistr} and \fig{fig_avastat} exhibit several
interesting features. The size of the typical avalanche quickly
increases with increasing liquid content reflecting the increasing
cohesion, then a decrease is observable when the viscoplastic
regime is reached. At the lowest liquid contents we observe single
peaks, i.e. the system has a characteristic avalanche size with
noise-like variations. At larger liquid contents, however, in
several cases we observe double peaks or extremely broad avalanche
size distributions. This implies the existence of two or more
characteristic avalanche sizes, i.e. both small and large
avalanches may occur. A similar feature has been observed in fine
cohesive powders \cite{Viturro01}. As discussed below, this
phenomenon is related to the fact that at high liquid contents the
surface is far from being flat and both the limit of stability and
the size of the avalanche depend strongly on the shape of the
actual surface.

Within a rotating drum experiment, the initial condition for an
avalanche is the surface profile formed by the previous avalanche.
The shapes of the surfaces after successive avalanches can thus be
regarded as steps of a special dynamical map. In the granular
regime the freely rolling grains always smooth out the surface, so
this mapping is trivial. However, in the correlated regime, where
the existence of clumps leads to rugged surfaces, this map is
rather complex. Due to the inherent fluctuations in the system, in
most cases this mapping is stochastic, smaller and larger
avalanches follow each other in a largely random manner. In some
cases, however, the system reaches quasi-stable fixed points (when
all the avalanches are very similar and their initial and final
surfaces are rotated versions of each other), or limit cycles,
when the initial profile is recovered in two or more avalanches.
Two examples for this latter case are shown in
\fig{fig_alternate}. Unfortunately the fluctuations in our system
usually destroy regular behavior, and thus we could not perform a
detailed investigation of this map. But we expect that a
corresponding model system without strong fluctuations would
display an extremely rich chaotic behavior.

%%%%%%%%%%%%%%%%%%%%%%%%%%%%%%%%%%%%%%%%%%%%%%%%%%%%%%%%%%%%%%%%%%%%%%
%%%%                                                              %%%%
\begin{figure}
 \includegraphics[clip,width=\columnwidth]{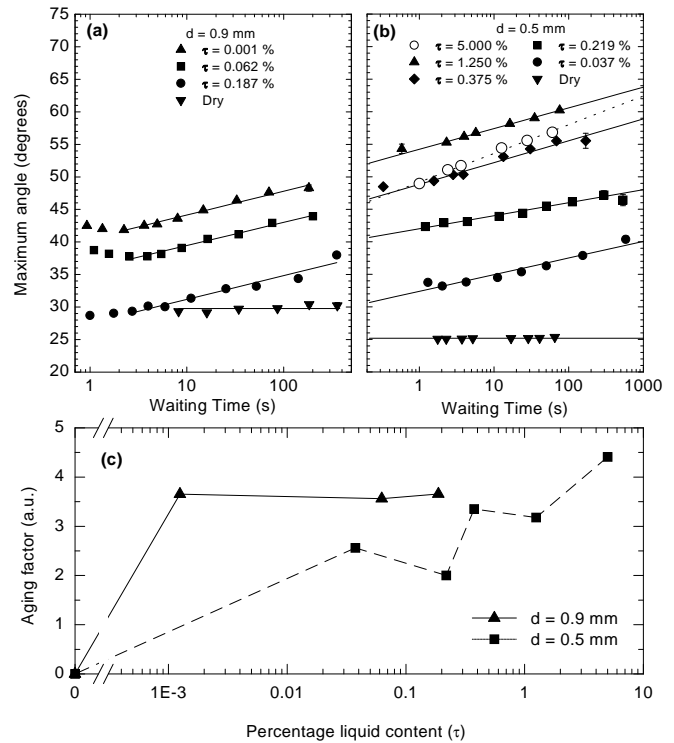}
\caption{\fpc{a-b} The dependence of the maximum angle $\tmax$ on
the waiting time between successive avalanches. The system
exhibits aging: longer waiting times between avalanches leads to
increased stability. The continuous lines are logarithmic fits to
the data. \fpc{a} \bead{0.9}, \fpc{b} \bead{0.5}. \fpc{c} The
intensity of aging (i.e. the slope of the fitted straight lines in
\fp{a} and \fp{b}) as a function of liquid content for the two
different bead sizes.}
 \label{fig_drum_aging}
\end{figure}
%%%%                                                              %%%%
%%%%%%%%%%%%%%%%%%%%%%%%%%%%%%%%%%%%%%%%%%%%%%%%%%%%%%%%%%%%%%%%%%%%%%

\subsection{Logarithmic aging}

As expected from previous studies, and shown in Section
\ref{sec_maxmin}, $\tmax$ strongly depends on the liquid content
due to the increasing capillary forces. Our measurements indicate
that $\tmax$ also depends on the rotation rate, $\Omega$. Although
(for small $\Omega$) the rotation rate only influences the waiting
time between successive avalanches, $\tmax$ clearly decreases with
increasing rotation rate. This effect (which is also visible in
the measurements of the surface as a function of drum rotation rate shown below) is presented in \fig{fig_drum_aging}\fp{a}
and \fp{b}, which show $\tmax$ as a function of the waiting time,
i.e. the time between avalanche events which decreases with
increasing rotation rate. The curves indicate a logarithmic
increase in the stability of the system. In \fig{fig_drum_aging}
\fp{c} we also present the intensity of the aging process
(quantified as the slope of the fitted straight lines). We do not
observe aging for the dry samples, and the effect is somewhat
stronger for the larger liquid contents, though on the basis of
our data we cannot determine the exact functional form.

\begin{figure}
 \vglue0.8cm
 \includegraphics[clip,width=\columnwidth]{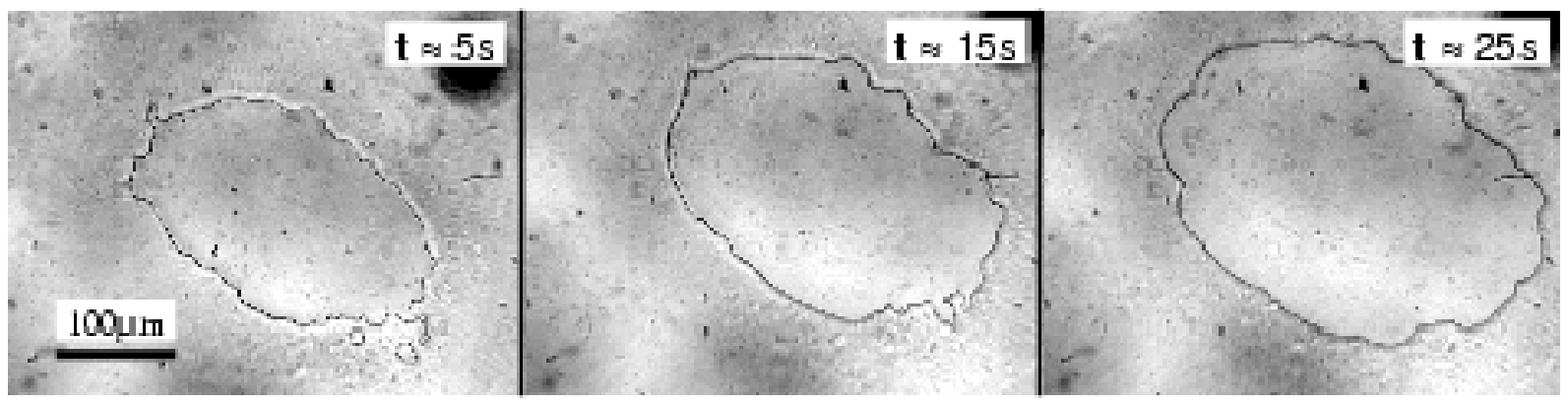}
 \vglue0.8cm
\caption{\redfig The oil at the contact point of a wet bead (similar to
the ones used in the experiment, but \bead{1.2}) and the glass
plate on which it rests.  The images are taken from above at
different time instants via phase-contrast microscopy. The delay
time from the instant when the bead was placed on the plate was
\fpc{a} $t\approx5\,$s, \fpc{b} $t\approx15\,$s, \fpc{c}
$t\approx25\,$s. The flow of oil to the contact point thus occurs
on a timescale of several seconds, suggesting a natural
explanation for our observed aging effects.
 }
 \label{fig_drum_oilatcontact}
\end{figure}

Similar aging effects have been observed in several different
granular experiments \cite{Bocquet98,Losert00}, but the underlying
mechanisms are complex, since several factors may play an
important role \cite{Losert00}. Since in our experiment the effect
is more pronounced at higher liquid contents, and the vapor
pressure of oil is low, we attribute it to the motion of oil
flowing towards the contact points rather than condensation
effects \cite{Bocquet98}. This assumption is also supported by our
investigations of a wet contact by phase contrast microscopy. We
placed a wet bead on the object plate and observed a flow of oil
to the region of the contact point (\fig{fig_drum_oilatcontact}).
Our observations revealed that a liquid bridge was formed
instantaneously when the bead was placed on the plate, but the bridge's shape continued changing slowly even after several seconds of waiting
time. This finding further supports the importance of viscosity to
the stability of wet granular media which was demonstrated by
Samandani et al. \cite{Samadani00,Samadani01}.

%%%%%%%%%%%%%%%%%%%%%%%%%%%%%%%%%%%%%%%%%%%%%%%%%%%%%%%%%%%%%%%%%%%%%%
%%%%                                                              %%%%
%%%%                                                              %%%%
%%%%%%%%%%%%%%%%%%%%%%%%%%%%%%%%%%%%%%%%%%%%%%%%%%%%%%%%%%%%%%%%%%%%%%

\section{The role of the rotation rate: avalanches and continuous flow}
\label{sec_phdiag}

%%%%%%%%%%%%%%%%%%%%%%%%%%%%%%%%%%%%%%%%%%%%%%%%%%%%%%%%%%%%%%%%%%%%%%
%%%%                                                              %%%%
\begin{figure}
 \includegraphics[clip,width=7cm]{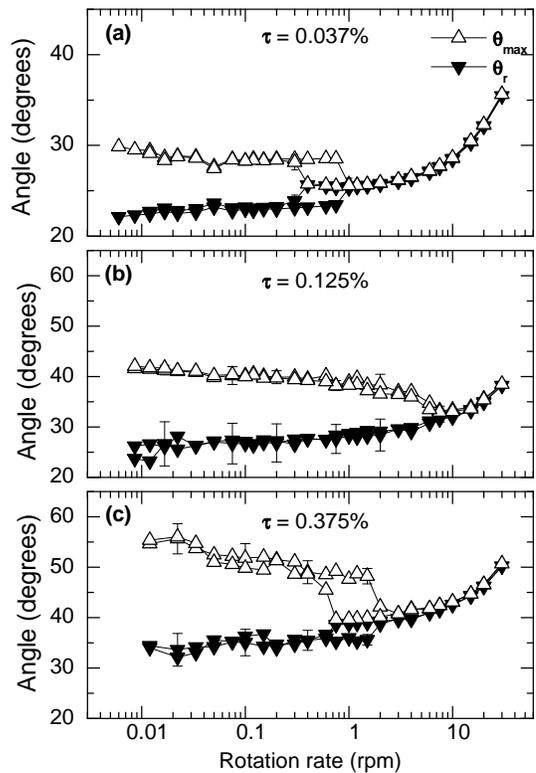}%
\caption{ \fpc{a} \fpc{b} \fpc{c} $\tmax$ and $\tr$ as a function of rotation
rate for 3 different liquid contents, \bead{0.5}. The error bars
represent the standard deviation of the observed angle
distributions. } \label{fig_hyst}
\end{figure}
%%%%                                                              %%%%
%%%%%%%%%%%%%%%%%%%%%%%%%%%%%%%%%%%%%%%%%%%%%%%%%%%%%%%%%%%%%%%%%%%%%%

\subsection{Hysteresis in the transition between continuous and avalanching flow}

As mentioned above, if the rotation rate is increased above a
threshold the system exhibits continuous flow instead of discrete
avalanches. Earlier studies in dry media revealed that the
transition between avalanching and continuous flow is hysteretic
in rotation rate \cite{Rajchenbach90,Caponeri95}. Now we have
investigated the avalanching -- continuous flow transition for
various liquid contents. We slowly increased and then decreased
the rotation rate and measured the average surface angle in the
continuous regime as well as $\tmax$ and $\tr$ in the avalanching
regime. The rotation rate was changed in discrete steps by factors
between $1.2$ and $1.4$, each step was performed with a constant
angular acceleration in $3$ seconds. Between the steps the drum
was rotated for typically $0.5-1$ full rotation to determine
whether the flow is continuous or avalanching (the minimum
rotation was $90^\circ$ at some of the lowest rotation rates, the
maximum was $40$ full rotations at \rot{30}). We consider the flow
continuous when the medium never comes to rest with respect to the
drum, which coincides with $\Delta\theta\equiv\tmax-\tr<1^\circ$.
The results of typical runs are presented in \fig{fig_hyst}\fp{a},
\fp{b} and \fp{c}, where $\tmax$ and $\tr$ are plotted as a function of
the rotation rate.

The three basic regimes of behavior observed previously are also
reflected in the nature of the transition. In the granular regime
at low liquid contents, (e.g. \oil{0.037}), where the behavior is
qualitatively similar to the dry case, we observe a clear
hysteretic transition between continuous flow and avalanching
(\fig{fig_hyst}\fp{a}). At a somewhat higher liquid content
(\oil{0.125}, see \fig{fig_hyst}\fp{b}) the correlated behavior is
marked by a lack of hysteresis, and intermittent avalanches are
observed at a relatively high rotation rate compared to the other
regimes. The continuous flow in this regime consists of a stream
of separated clumps rather than the constant flux seen in the
other two regimes.  At \oil{0.375} (\fig{fig_hyst}\fp{c})
hysteresis is again observed, coinciding with the onset of the
viscoplastic continuous flow, which is smooth and coherent over
the entire sample.

%%%%%%%%%%%%%%%%%%%%%%%%%%%%%%%%%%%%%%%%%%%%%%%%%%%%%%%%%%%%%%%%%%%%%%
%%%%                                                              %%%%
\begin{figure}[!tbp]
 \includegraphics[clip,width=\columnwidth]{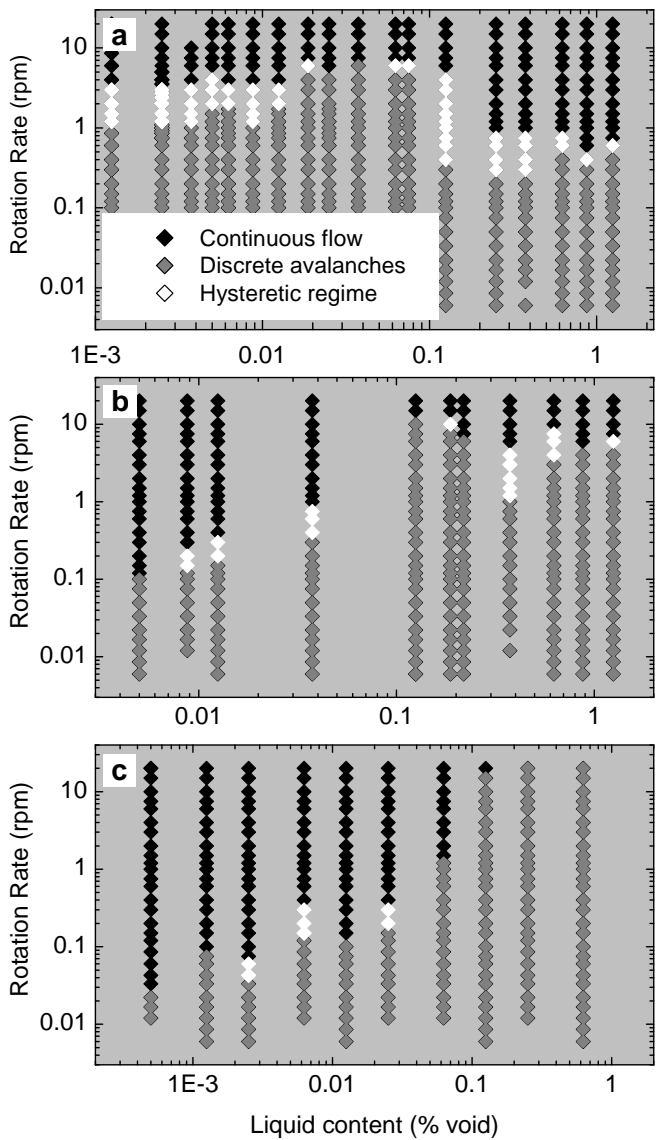}
 \caption{\fpc{a} Phase diagram of the dynamical behavior as a function of the
liquid content ($\tau$) and the rotation rate ($\Omega$). \fpc{a}
large beads, \bead{0.9}, \fpc{b} medium beads, \bead{0.5}, \fpc{c}
small beads, \bead{0.35}.}
 \label{fig_phdiag}
\end{figure}
%%%%                                                              %%%%
%%%%%%%%%%%%%%%%%%%%%%%%%%%%%%%%%%%%%%%%%%%%%%%%%%%%%%%%%%%%%%%%%%%%%%

The origin of the hysteresis in the transition between continuous
and avalanching flow is a fascinating subject 
\cite{Jaeger89,Rajchenbach90,Caponeri95,Benza93}, but a detailed
discussion goes beyond the scope of this paper. We hypothesize
that the hysteresis is observed because the transitions on
increasing and decreasing rotation rate have different origins,
thus they do not necessarily coincide. A rough explanation of the
transition from avalanching to continuous flow is that the
successive avalanches merge. More precisely, at high velocities
the rotation during the avalanche is becoming important and the
constant energy supply prevents the grains from stopping. For the
reverse transition, the continuous flow cannot be maintained at
arbitrarily low velocities, because then the flowing layer cannot
stay fluidized, and the grains get stuck in small depressions of
the underlying layer.

The nature of the transition is naturally different in the different wetting regimes. In the granular and the
viscoplastic regime the constant material flux in the continuous
flow phase is qualitatively different from the first increasing
and then decreasing flux observed in the case of an avalanche.
Thus the intermittent and the continuous flow may represent two
locally stable dynamical phases with a hysteretic transition
between them. On the other hand in the correlated regime there is
no qualitative change in the flow dynamics when the motion becomes
continuous: the successive clumps do not merge even if their
motion overlaps in time. According to the arguments advanced by
Rajchenbach \cite{Rajchenbach90}, the characteristic fall time of
the grains is thus the same in the avalanching and the continuous
flow. Therefore we observe no hysteresis.

\subsection{The phase diagram }

We summarize our measurements of the different flow regimes in a
phase diagram (\fig{fig_phdiag}) indicating the nature of the flow
as a function of liquid content and rotation rate. Each symbol
corresponds to a measurement, grey symbols denote avalanches,
black symbols denote continuous flow, white symbols mark the
hysteretic regions. The most complete graph corresponds to the
large, \bead{0.9}, beads (\fig{fig_phdiag}\fp{a}). Here the three
regimes are clearly visible. Near $\tau\approx 0.02$\,\% the
disappearance of hysteresis and the displacement of the transition
point towards higher rotation rates indicates the transition from
the granular to the correlated regime. The sudden change in the
transition point around $\tau\approx 0.1$\,\% marks the onset of
the viscoplastic continuous flow.

The other two diagrams (\fig{fig_phdiag}\fp{b} and \fp{c}, medium
(\bead{0.5}) and small (\bead{0.35}) beads, respectively) contain
fewer points, but the main trends are visible. A universal feature
of the phase diagrams is that, upon approaching the correlated regime
from the granular one the transition point moves towards higher
rotation rates.  This suggests that the clump formation leads to more effective
transport and quicker avalanches as was noted in our previous
draining crater studies \cite{Tegzes99}. A general trend is that
for smaller beads the transition to correlated behavior occurs at
higher liquid contents, most probably due to the larger internal
surface of these samples.

The most striking feature is that, as the bead size is decreased,
the viscoplastic behavior and in particular the region of the
viscoplastic continuous flow diminishes. For the \bead{0.5} beads
it is restricted to the region between
$0.3\,\%\lesssim\tau\lesssim0.8\,\%$ and
$1$\,rpm$\lesssim\Omega\lesssim 10$\,rpm, and the \bead{0.35}
beads exhibit no viscoplastic flow at all. A possible explanation
is that, in samples consisting of smaller beads, the number of
contacts at the boundary of a given volume of material is much
larger, therefore the cohesive forces dominate inertial effects.
For smaller beads the hysteresis of the continuous-to-avalanche
transition also seems to decrease, but this is difficult to
interpret and will require further investigation to understand.

We have performed some preliminary measurements on the role of the
viscosity of the interstitial liquid. The results (not shown here)
indicated that the fluid viscosity is indeed an important
parameter, e.g. experiments with a highly viscous silicon oil
yielded viscoplastic flow in a wider range of liquid contents and
rotation rates.  These results are also consistent with the
findings of Samandani and Kudrolli \cite{Samadani01}.

\section{Surface morphology}
\label{sec_morph}

\begin{figure}
 \includegraphics[clip,width=\columnwidth]{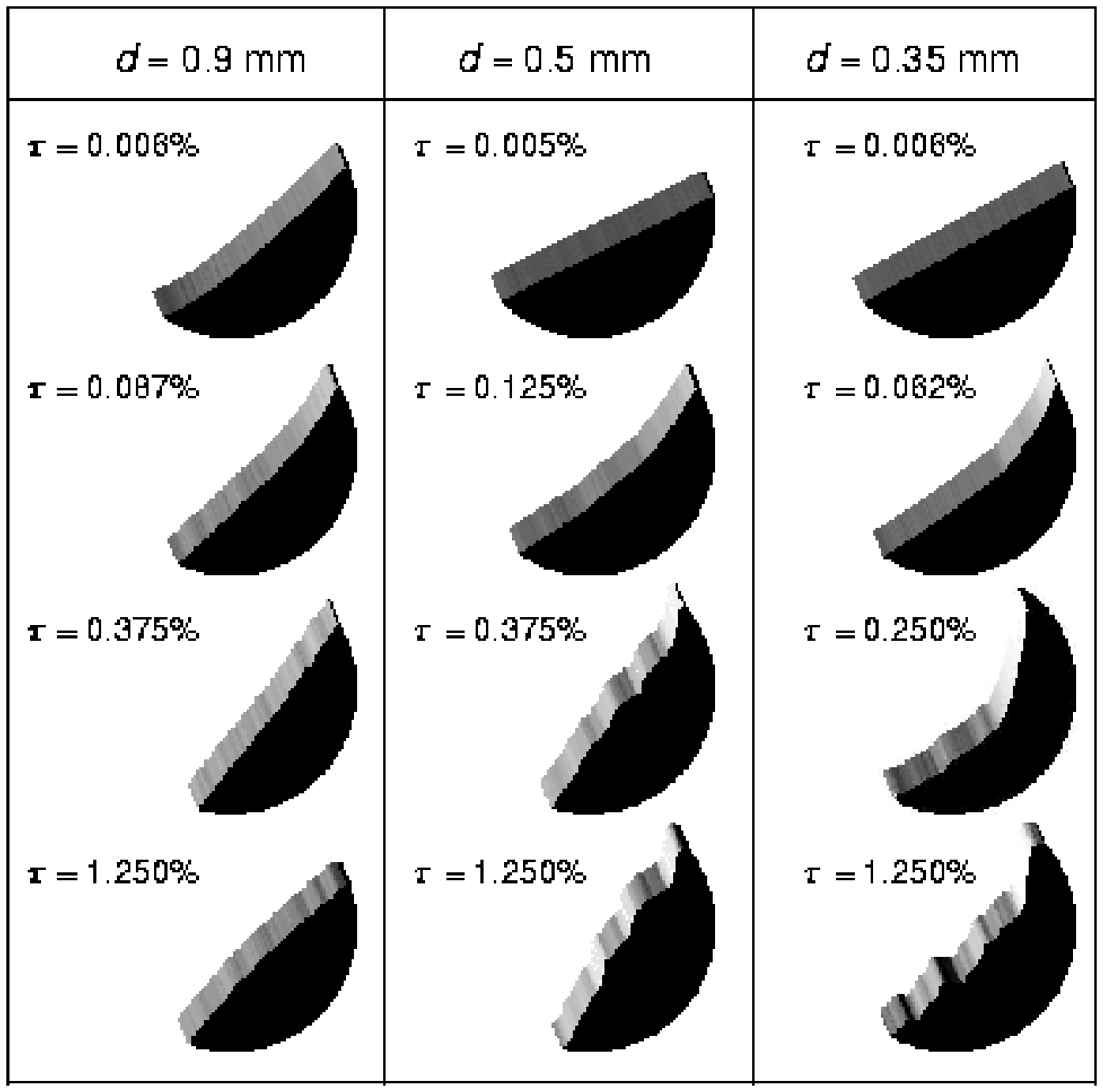}
\caption{\redfig Sample surfaces for various materials, as indicated in
the figure. The snapshots were taken right before an avalanche,
the rotation rate was \rot{0.05} for the large and medium beads
and \rot{0.043} for the small beads. Note that the surface is
close to flat in the granular regime, mostly concave in the
correlated regime and mostly convex in the viscoplastic regime.
Note also the smaller scale structure at the smaller bead sizes. }
 \label{fig_shape_allbeads}
\end{figure}

Up to this point we have described the surface by a single
quantity, the overall surface angle, but the actual surfaces are
rough and have significant curvature. This morphology contains
additional information about the system, and should be useful in
producing realistic models of wet granular flow.
\fig{fig_shape_allbeads} shows sample surface profiles for various
materials. The graph confirms that the wet samples form exotic
surfaces, especially for the smaller beads. Note that, besides the
overall shape of the surface, some smaller scale roughness is also
observable for the larger liquid contents. These small scale
features typically are random and fluctuate between avalanches. In
the viscoplastic regime, however the coherence of the flow
strongly reduces fluctuations, and a robust pattern is formed that
is reproduced at the end of each avalanche (as discussed in
section \ref{sec_pattern} below).

%%%%%%%%%%%%%%%%%%%%%%%%%%%%%%%%%%%%%%%%%%%%%%%%%%%%%%%%%%%%%%%%%%%%%%
%%%%                                                              %%%%
%%%%                                                              %%%%
%%%%%%%%%%%%%%%%%%%%%%%%%%%%%%%%%%%%%%%%%%%%%%%%%%%%%%%%%%%%%%%%%%%%%%

%%%%%%%%%%%%%%%%%%%%%%%%%%%%%%%%%%%%%%%%%%%%%%%%%%%%%%%%%%%%%%%%%%%%%%
%%%%                                                              %%%%
\begin{figure}
 \includegraphics[clip,width=7.7cm]{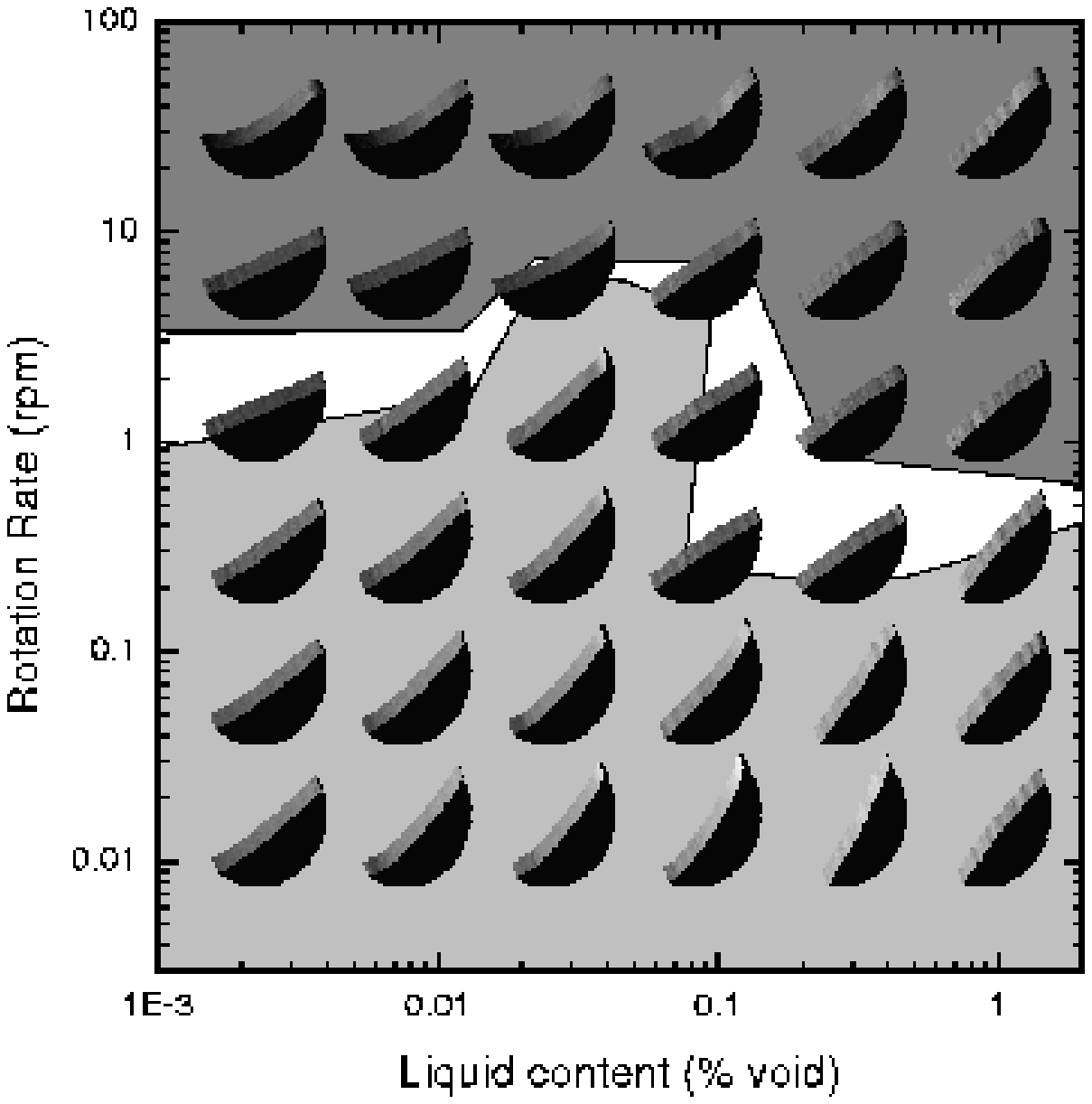}
\caption{\redfig Shape of the surface at various points of the parameter
space for the large, \bead{0.9} beads. A sketch of the phase
diagram is shown in greyscale in the background: the dark grey
region denotes continuous flow, the light grey region corresponds
to avalanches; the white area is the hysteretic regime. The shape
changes both as a function of $\Omega$ and $\tau$. At the highest
rotation rates the S-shaped regime \cite{Rajchenbach90} is clearly
observable.
 }
 \label{fig_drum_shape}
\end{figure}
%%%%                                                              %%%%
%%%%%%%%%%%%%%%%%%%%%%%%%%%%%%%%%%%%%%%%%%%%%%%%%%%%%%%%%%%%%%%%%%%%%%

%%%%%%%%%%%%%%%%%%%%%%%%%%%%%%%%%%%%%%%%%%%%%%%%%%%%%%%%%%%%%%%%%%%%%%
%%%%                                                              %%%%
\begin{figure}
\vglue 0.5cm
 \includegraphics[clip,width=\columnwidth]{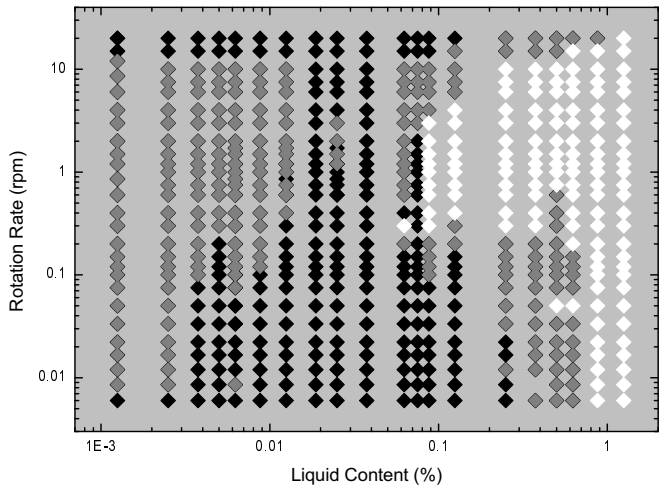}
\caption{The value of convexity parameter $\kappa$ as a function
of the rotation rate and the liquid content for the large beads,
\bead{0.9}. Black symbols: concave surface ($\kappa<0$), white
symbols: convex surface ($\kappa>0$), grey symbols:
($\kappa\approx0$).
 }
 \label{fig_drum_convex}
\end{figure}
%%%%                                                              %%%%
%%%%%%%%%%%%%%%%%%%%%%%%%%%%%%%%%%%%%%%%%%%%%%%%%%%%%%%%%%%%%%%%%%%%%%

%%%%%%%%%%%%%%%%%%%%%%%%%%%%%%%%%%%%%%%%%%%%%%%%%%%%%%%%%%%%%%%%%%%%%%
%%%%                                                              %%%%
\begin{figure}
 \includegraphics[clip,width=\columnwidth]{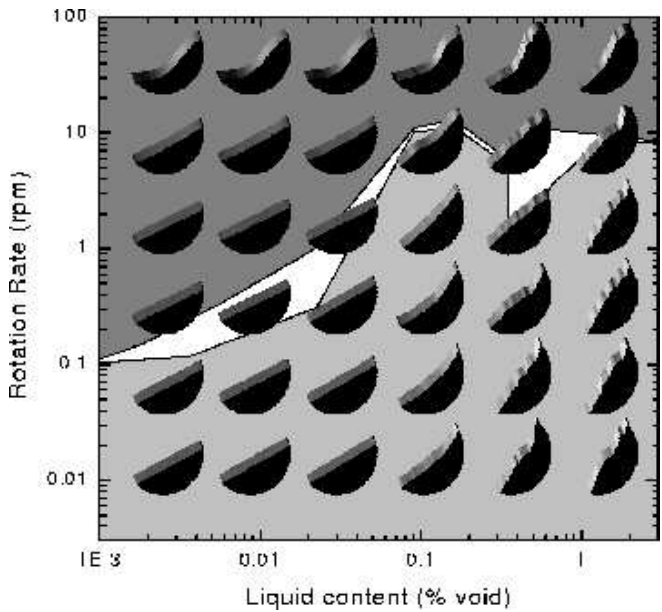}
\caption{\redfig Shape of the surface at various points of the parameter
space, for the medium, \bead{0.5} beads (see the caption of
\protect\fig{fig_drum_shape} for further explanation). The
observed surface shapes are more complicated for these
beads compared to the \bead{0.9} beads, but the overall convex vs. concave nature is similar.}
 \label{fig_drum_shape_small}
\end{figure}
%%%%                                                              %%%%
%%%%%%%%%%%%%%%%%%%%%%%%%%%%%%%%%%%%%%%%%%%%%%%%%%%%%%%%%%%%%%%%%%%%%%

Since the surfaces of the largest (\bead{0.9}) bead samples are
relatively simple, we investigate them in more detail.
\fig{fig_drum_shape} shows sample surfaces for the \bead{0.9}
beads at various liquid contents and rotation rates. In the
granular regime the surface is almost flat, correlated avalanches
typically form concave surfaces, and the viscoplastic flow
typically results in a convex surface. Furthermore, at the highest
rotation rates, the surface is S-shaped due to inertial effects
across the entire wetness range \cite{Rajchenbach90}. The
convexity of the surface can be characterized quantitatively
several ways \cite{Smith01}.  A simple and robust measure is the integrated area ($\kappa$)
between the surface profile ($h(x)$) and the straight line
connecting the two ends of the surface line. For a flat surface
where $h(x)$ is a straight line, $\kappa=0$; $\kappa<0$ implies a
predominantly concave surface, $\kappa>0$ implies a predominantly
convex surface.

In \fig{fig_drum_convex} the value of $\kappa$ is indicated by the
color of symbols as a function of $\tau$ and $\Omega$. Black
symbols correspond to concave surfaces (negative $\kappa$), and
approximately indicate the correlated regime. White symbols are
used for positive $\kappa$ values, corresponding to
the viscoplastic regime. Grey symbols indicate that
$\kappa\approx0$, either because the surface is flat (upper left
region of \fig{fig_drum_convex} - this is approximately the
granular regime), or because convex and concave parts are balanced
(at the boundary of correlated and viscoplastic regimes).  For comparison we also show the typical observed surface shapes
for the medium size (\bead{0.5}) beads
(\fig{fig_drum_shape_small}). Also here the convexity of the
surface seems to correlate with the different flow regimes,
however, some additional smaller scale features are also
observable. 

Clearly, the phase boundaries in \fig{fig_drum_convex} are not
vertical lines. This means that the liquid content is not the only
parameter influencing the behavior: changing the rotation rate can
also switch between different flow mechanisms. This observation
implies that the formation of correlated clumps is a dynamical
process with characteristic times comparable to the other
timescales of the experiment. We expect that the different convexity in
the different regimes could be reproduced relatively easily in the framework of a continuum model. A detailed explanation of the surface shapes, however, would probably require
a full analysis of the avalanche dynamics.

\section{Detailed dynamics of avalanches}
\label{sec_avadyn}

\begin{figure}[tbp!]
 \includegraphics[clip,width=\columnwidth]{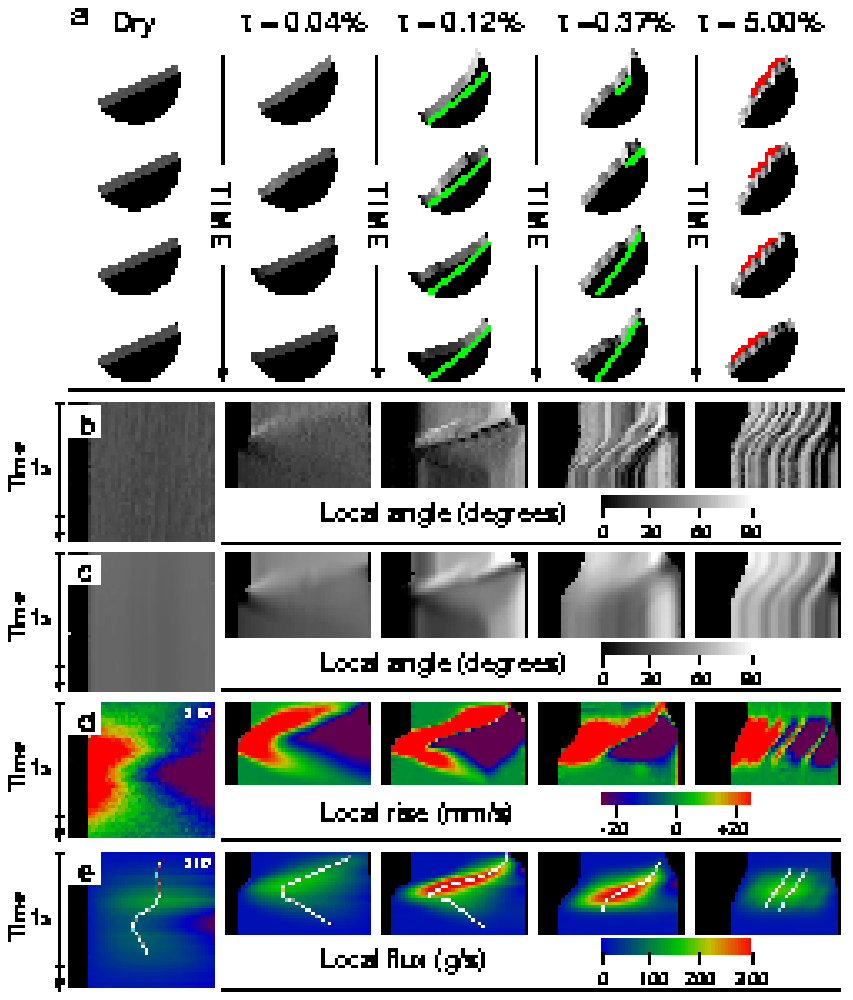}
 \caption{\redfig (color) Dynamics of avalanches of different types with grain
size \bead{0.5}, at \rot{0.12}. \fpc{a} Snapshots (at $0.1-0.2$\,s
time intervals) of five single avalanches corresponding to
different liquid contents. The third dimension is used for a
brightness-coded representation of the local slope ($\alpha$). The
green lines indicate approximate slip planes in the correlated
regime, and the red line shows the traveling quasi-periodic
surface features in the viscoplastic regime. \fpc{b} The local
slope with the same brightness coding, as a function of space and
time. A horizontal line corresponds to a surface profile at a
given instant, and time increases downwards (thus avalanches
propagate down and to the left). The slanted bright and dark
regions correspond to the avalanche front and the kink
respectively (see text). The stripes at higher liquid contents
indicate lasting surface features. \fpc{c} \fpc{d} \fpc{e} The
characteristic features of the avalanches averaged over $300-500$
avalanche events. The displayed quantities are \fpc{c} the local
surface angle, \fpc{d} the rate of change of local height, and
\fpc{e} the local grain flux (the white lines indicate the point
of maximum current), respectively. Note that the surface patterns
at the highest oil contents are robust against averaging. }
\label{fig_drum_avadyn1}
\end{figure}

The rotating drum apparatus allows us to obtain information not
just about the medium before and after the avalanche events, but
also about the details of the grain motion {\em during avalanche
events}.  In order to analyze the dynamics of avalanches we have
obtained two-dimensional space-time matrices, $h(x,t)$,
characterizing the sample surfaces throughout the avalanche
process which can then be analyzed to produce a variety of
information about the individual avalanches.  By taking
derivatives of the $h(x,t)$ data, we obtain the local angle,
$\alpha (x,t)=\arctan [\partial _x h(x,t)$], and the local
vertical velocity, $u(x,t)=\partial _t h(x,t)$, of the surface
profile. Furthermore, by then integrating the vertical velocity
(using the continuity equation and assuming constant density
\cite{Douady99}), we also obtain the local flux in the avalanche,
i.e. $\phi (x,t) = \rho w \int_{-D/2}^x \partial_t h(x',t) dx'$
(where $\rho$ is the grain density, $w$ is the width of the drum,
and $D$ is the drum diameter), which represents the material
flowing through a given vertical plane at position $x$. In
\fig{fig_drum_avadyn1}\fp{a} we present snapshots of the progression
of single avalanches for several typical liquid contents.
\fig{fig_drum_avadyn1}\fp{b} displays $\alpha(x,t)$ as a function of
space (horizontally) and time (downwards) for the same individual
avalanches. In \fig{fig_drum_avadyn1} (c-e), we present the
average behavior of $300-500$ avalanches at the same liquid
contents, and show similar graphs of the time evolution of
$\avg{\alpha(x,t)}$, $\avg{u(x,t)}$, and $\avg{\phi(x,t)}$, where
$\avg{}$ denotes averaging over avalanches. By obtaining these
quantitative measures of the averaged properties, we can separate
the robust characteristics of the avalanche dynamics from the
large fluctuations which are inherent in avalanche processes.

%%%%%%%%%%%%%%%%%%%%%%%%%%%%%%%%%%%%%%%%%%%%%%%%%%%%%%%%%%%%%%%%%%%%%%
%%%%                                                              %%%%
%%%%                                                              %%%%
%%%%%%%%%%%%%%%%%%%%%%%%%%%%%%%%%%%%%%%%%%%%%%%%%%%%%%%%%%%%%%%%%%%%%%

The averaging process raises a few technical questions. First, the order
of averaging and numerical differentiation is optional, i.e. we can either
calculate $\alpha(x,t)$ and the other quantities from the raw $h(x,t)$ data
and perform averaging afterwards, or take the averaged $\avg{h(x,t)}$ and
calculate its numerical derivatives. We verified that the two methods yield
equivalent results, and chose the former option.

A second question is how to find the corresponding frames in
different recorded avalanches. In order to make this question
clear let us consider the averaging in more detail. Let us suppose
that we have recorded the variation of the surface profile for $N$
avalanches, and we have $N$ matrices $\huv^{(i)}$, $i=1..N$
denoting the local height in the $i$th avalanche at the position
$x=-D/2+u\,\delta x$ and time $t=t_0^{(i)}+v\,\delta t$, where
$\delta x$ and $\delta t$ are the spatial and temporal resolutions
respectively, and $t_0^{(i)}$ is ``the starting time'' of the
$i$th avalanche. Then, to obtain the average behaviour, we calculate
$\avg{\huv}=\sum_{i=1}^{N}\huv^{(i)}/N$.  Since the
properties of the avalanches (e.g. duration) vary between avalanche events, the appropriate choice for $t_0^{(i)}$ is not
trivial. We tried several algorithms and then decided to use the
most robust one which was based on the variation of the overall
surface angle $\theta(t)$. We fitted straight lines to the
segments of the sawtooth signal of $\theta(t)$ (see
\fig{fig_ang_raw}), and calculated the time $t_{int}^{(i)}$
corresponding to the intersection of the straight lines at the
beginning of an avalanche. Then we chose
$t_0^{(i)}=t_{int}^{(i)}-0.5$\,sec.

A third problem arises from the fact that the $\huv^{(i)}$
matrices have empty elements: the surfaces do not extend from
$-D/2$ to $D/2$. Simply excluding the empty elements from the
averaging leads to the appearance of some artifacts at the edges.
To resolve this issue, we filled in the empty elements with the
value of the closest elements in that row (as if the surface
continued horizontally over the edge of the drum), performed the
averaging, and then cut down those parts of the averaged profile
that were outside the drum. Note that this procedure only
influenced the behavior very close to the edges of the sample.

%%%%%%%%%%%%%%%%%%%%%%%%%%%%%%%%%%%%%%%%%%%%%%%%%%%%%%%%%%%%%%%%%%%%%%
%%%%                                                              %%%%
\begin{figure*}[!tbp]
 \includegraphics[clip,width=17cm]{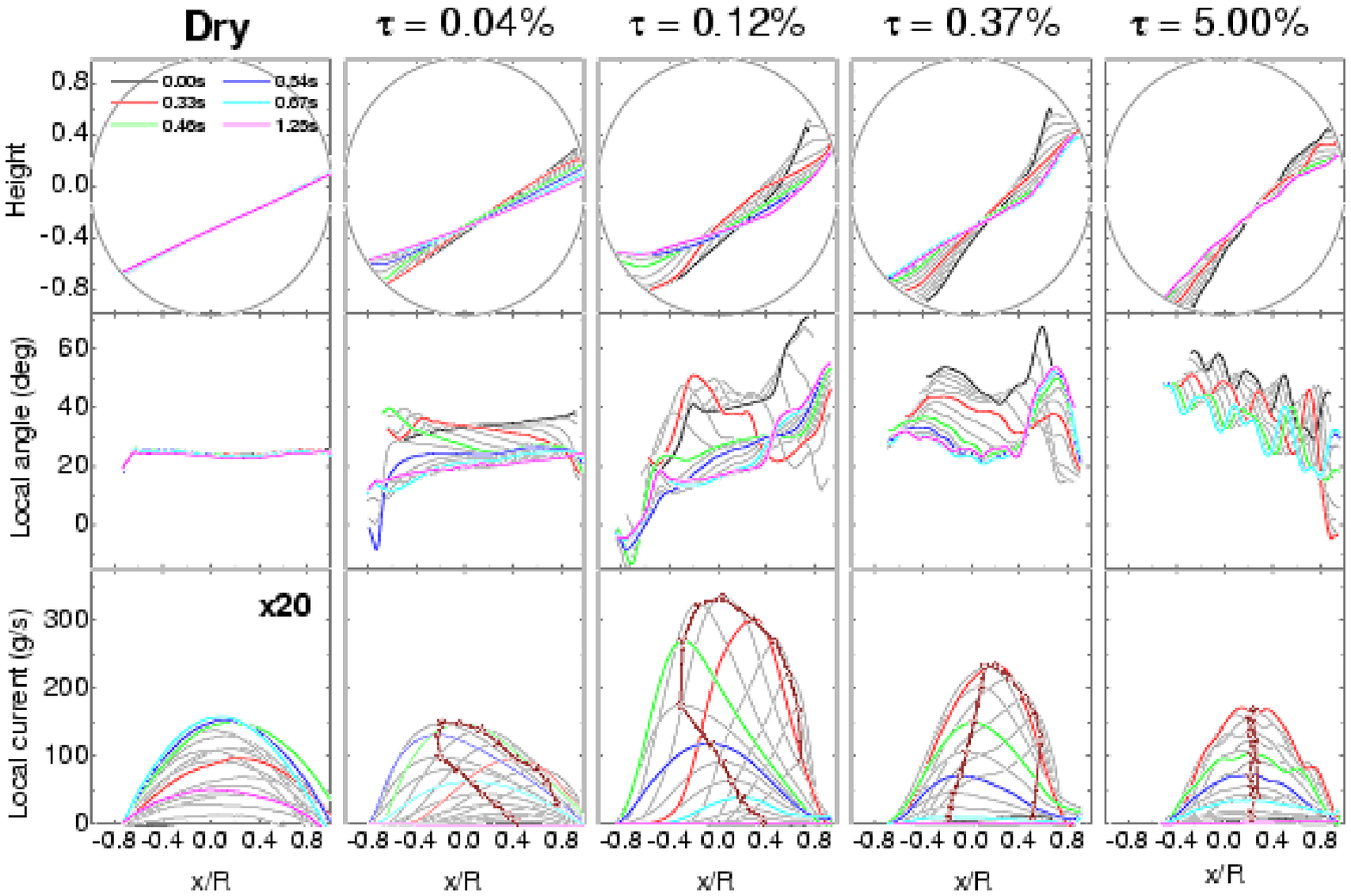}
\caption{\redfig (color) Averaged dynamical parameters: $\avg{h(x)}$,
$\avg{\alpha(x)}$, and $\avg{\phi(x)}$ as a function of space at
various time instants during the avalanche for 5 different liquid
contents, \bead{0.5}, \rot{0.12}. The time interval between
successive curves is $0.066$\,sec, some curves are marked with
color as specified in the legend. In the lowest row, triangles
mark the point of maximum current. The most important features for
different oil contents are the following. {\em Dry samples.} The
avalanches are very small, and our resolution is insufficient to
resolve the detailed dynamics. \oil{0.04} {\em Granular regime}.
The surface is close to flat, but the local angle is somewhat
larger near the top of the slope. The dynamics is dominated by fronts of
rolling grains. The region of steeper slope corresponding to the
downward propagating front as well as the negative slope of the
kink is visible in the graph of the local angle. The point of
maximum current also moves downhill then uphill due to the fronts.
\oil{0.12} {\em Correlated regime}: failure occurs along a slip
plane and a block slides down. The block is best seen in the red
curve, the edges are smoothed by the averaging over several
hundred avalanches. Some rolling grains are still present, as
marked by the presence of the kink. \oil{0.37} {Correlated
regime}, usually with multiple avalanches, but these are averaged
out. The fronts have disappeared,  and the angle near the bottom
of the slope decreases monotonically. There are lasting contacts
during the avalanche, thus the grains cannot roll, and the
material stops coherently as a block when it reaches the bottom.
\oil{5.00} {\em Viscoplastic regime.} The surface moves
coherently. The current extends over the whole surface during the
whole duration of the avalanche. Curves of the local angle reveal
the robust surface patterns as a series of local maxima. During
the avalanche the local maxima move downwards (and spread out
slightly), and a new maximum is formed near the top.
 }
 \label{fig_drum_xcurves}
\end{figure*}
%%%%                                                              %%%%
%%%%%%%%%%%%%%%%%%%%%%%%%%%%%%%%%%%%%%%%%%%%%%%%%%%%%%%%%%%%%%%%%%%%%%

In \fig{fig_drum_xcurves} we present the same data as series of
curves, where the columns again represent different liquid
contents. In the first row of graphs we plot the averaged surface
profiles ($\avg{h(x)}$) at different time instants during the
avalanche. The second row of graphs shows the variation of the
local angle ($\avg{\alpha(x)}$), and the third row plots the local
flux ($\avg{\phi(x)}$) with triangles showing the position of
maximum flux.   In the following we analyze the dynamics in the 3
regimes based on Figs. \ref{fig_drum_avadyn1} and
\ref{fig_drum_xcurves}.   We have also magnified parts of
\fig{fig_drum_avadyn1} in \fig{fig_drum_fronts} to emphasize
particular features in the data.  For comparison we also show the
most important avalanche types for the \bead{0.9} beads in
\fig{fig_drum_avadyn2}. Apart from the absence of robust
large-scale patterns for the \bead{0.5} beads (discussed below),
the qualitative features of the avalanches are quite similar for
the two types of beads although the effects are less pronounced
for the \bead{0.9} beads.

\subsection{The avalanche types}

{\em Granular avalanches.} With dry grains
\cite{Rajchenbach90,Liu91,Jaeger89,Daerr99,Bretz92,Frette96}, the surface
remains almost flat throughout the avalanche, and the avalanches have a much
longer duration and much smaller flux than in the wet media - as is expected
due to the lack of cohesion. Our resolution is not sufficient to distinguish
any propagating front in this case, but our system is very similar to that
investigated by \cite{Rajchenbach02}, all the surface grains are close to the
limit of their stability. Thus we assume that the failure mechanism is similar:
the propagation of the front destabilizing the grains is quicker than the
material flow. This type of front is marked only by a very slight dilation of
the material, which we cannot detect.

\begin{figure}[!tbp]
 \includegraphics[clip,width=6.5cm]{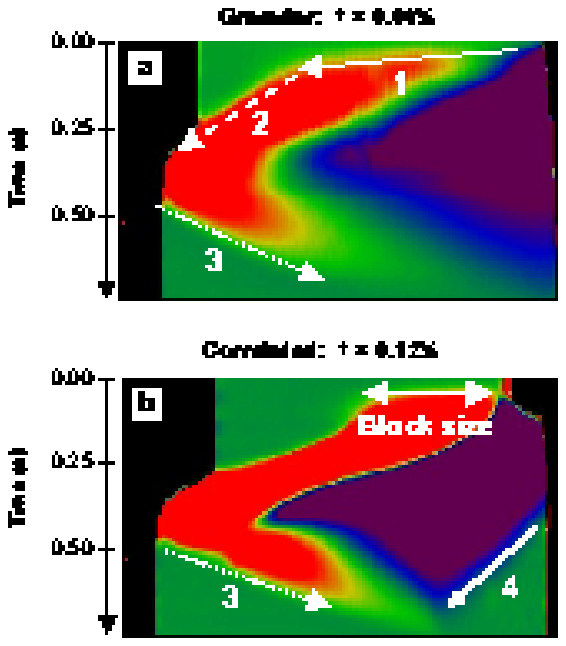}
\caption{\redfig (color) The various fronts observed in our setup in \fpc{a} the
granular regime (\oil{0.04}) and in \fpc{b} the low liquid content
region of the correlated regime (\oil{0.12}), for \bead{0.5},
\rot{0.12}.  The colors indicate the rate of change of local
height during the avalanche (see \protect\fig{fig_drum_avadyn1}).
We observe 4 different fronts. $1$: destabilizing front.
Propagates faster than the material flow via interactions between
neighboring grains
--- destabilizes particles. $2$: Avalanche front. Consists of
rolling grains, which proceed on a medium at rest and destabilize particles by
collisions. $3$: Kink. The rolling grains accumulate at the lower wall and form
a region of low slope there. This slows down the incoming grains, thus this
front travels uphill. $4$: Rear front. The moving grains leave the static ones
behind.
 }
 \label{fig_drum_fronts}
\end{figure}

With the addition of very small amounts of liquid (e.g. \oil{0.04}, see Fig. \fig{fig_drum_fronts}\fp{a}) the
avalanches become much larger due to the onset of intergrain
cohesion --- this allows us to observe the dynamics in more
detail. In this granular regime the avalanche is always initiated
at the top of the surface, and the upper part of the surface is
quickly destabilized. As the particles start moving downwards, a
front of rolling grains travels downhill. The grains in the lower
regions remain at rest until the rolling grains reach them. This
type of front corresponds to the ``start down'' front of Douady et
al.\ \cite{Douady01}. The difference in behavior compared to the
perfectly dry material can probably be explained by the slight
concave curvature of the surface: the grains lower on the slope are
in a relatively stable state when the avalanche is initiated.

When the downward-propagating front reaches the wall of the drum at the bottom
of the slope, a kind of shockwave is formed, and the rolling of grains is
stopped by a region of smaller local angle travelling uphill. The second front
is classified as a ``stop up'' front by \cite{Douady01}, and corresponds to the
``kink'' seen in other experiments on dry beads (generally associated with
individually rolling grains reaching a solid barrier at the bottom of the slope
\cite{Makse97,Gray98,Cizeau99}).

%%%%%%%%%%%%%%%%%%%%%%%%%%%%%%%%%%%%%%%%%%%%%%%%%%%%%%%%%%%%%%%%%%%%%%
%%%%                                                              %%%%
%%%%                                                              %%%%
%%%%%%%%%%%%%%%%%%%%%%%%%%%%%%%%%%%%%%%%%%%%%%%%%%%%%%%%%%%%%%%%%%%%%%

{\em Correlated Avalanches.} At intermediate liquid contents, in
the correlated regime ($0.1 \,\%<\tau < 2\,\%$), the principal
failure mechanism is a fracture along a curved slip plane
(approximated by the green lines in \fig{fig_drum_avadyn1}\fp{a}),
analogous to the dynamics of a class of geological events known as
``slides'' \cite{Dikau96}. This is the failure mechanism described
by the Mohr-Coulomb model \cite{Nedderman92,Halsey98}. At
\oil{0.12} there is a single slip plane, but at larger liquid
contents (\oil{0.37}) the avalanches occur through a succession of
local slip events. The medium becomes more cohesive with
increasing liquid content, preventing grains from moving
individually, and thus the kink disappears for $\tau \ge 0.3\,\%$
since the material moving as a connected block stops coherently
when it hits the bottom.

{\em Viscoplastic Avalanches.} The
onset of the viscoplastic regime ($\tau \approx2\,\%$) is
accompanied by dramatic changes in the behavior. The flow becomes
correlated across the entire granular surface as demonstrated by
the parallel lines during the avalanche in \fig{fig_drum_avadyn1}
\fp{b} and \fp{b}.  Since the whole surface moves coherently (rather
than breaking apart and evolving separately in different parts of
the surface), fluctuations are strongly suppressed
\cite{Tegzes99}. This behavior is qualitatively similar to a
different class of geological event, called a "debris flow" or
"mudflow" \cite{Dikau96,Iverson97,Coussot97}. The coherent nature
of the motion in this regime leads to novel aspects of the
dynamics which are not observed in other granular systems.

\subsection{Pattern formation in the viscoplastic regime}
\label{sec_pattern}

One novel property of the viscoplastic avalanches is the robust
topology of the top surface which spontaneously forms a nearly
periodic pattern (seen in \fig{fig_drum_avadyn1}\fp{a}). This
surface structure is maintained essentially intact during the
avalanche (note that the lines are continuous throughout the
avalanche in \fig{fig_drum_avadyn1}\fp{b}), indicating that there
are lasting contacts in the flowing layer. Moreover, the pattern
is not random, but rather has features which are reproduced at the
end of each avalanche.  This is demonstrated most clearly in
\fig{fig_drum_xcurves} and in \fig{fig_drum_avadyn1}\fp{b}, where
the average of $347$ avalanches of the \oil{5} sample has the same
features as the typical individual avalanche shown in
\fig{fig_drum_avadyn1}\fp{a} and \fp{b}. The robust nature of the
surface structure of the wettest grains is in sharp contrast to
the other regimes where averaging completely smoothes out the
smaller surface features. We can understand this behavior as
resulting from coherence of the entire flow, which strongly
reduces fluctuations in this regime. With minimal fluctuations,
the final surface structure after each avalanche is essentially
the same, thus setting the same initial condition for the next
avalanche.  With the same initial conditions for each avalanche,
naturally the surface features are reproduced each time.

\begin{figure}[!tbp]
 \includegraphics[clip,width=\columnwidth]{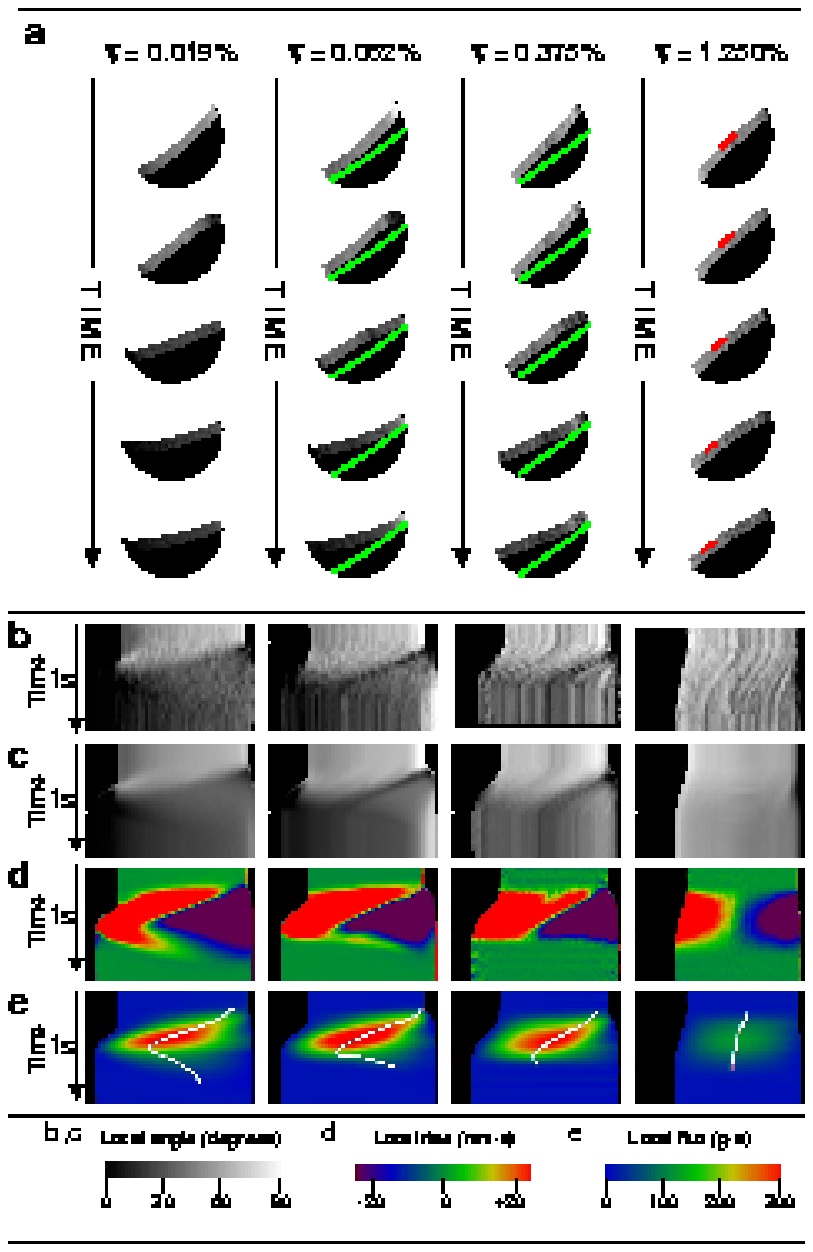}
\caption{\redfig (color) Avalanche dynamics in samples consisting of large beads
\protect\bead{0.9}. (See also the caption of
\protect\fig{fig_drum_avadyn1}.)}
 \label{fig_drum_avadyn2}
\end{figure}

Our experiments with the larger (\bead{0.9}) beads also reveal
some pattern formation (see \fig{fig_drum_avadyn2}\fp{b}), but
with a smaller characteristic size corresponding to $8-10$ grain
diameters. The difference is probably due to the smaller ratio of
the cohesive forces to the gravitational forces on the grains. Due
to the small length scale of these patterns, they are largely averaged out, but we have also observed robust patterns that are repeated
after each avalanche under several different sets of conditions
(see e.g. \fig{fig_drum_avadyn2}, \oil{0.375})  (these other patterns are not as periodic as  at \oil{5}, \bead{0.5}). 
The dependence of the pattern on grain size and other factors will
be the subject of a future investigation.

%%%%%%%%%%%%%%%%%%%%%%%%%%%%%%%%%%%%%%%%%%%%%%%%%%%%%%%%%%%%%%%%%%%%%%
%%%%                                                              %%%%
%%%%                                                              %%%%
%%%%%%%%%%%%%%%%%%%%%%%%%%%%%%%%%%%%%%%%%%%%%%%%%%%%%%%%%%%%%%%%%%%%%%

\section{Dynamic properties of continuous flows}
\label{sec_flowdyn}

In the previous section we demonstrated that the avalanche dynamics
are dramatically different in the three regimes of behavior.  The different dynamics are also evident at higher rotation rates where the flow is
continuous. In this section we compare the properties of the
continuous flow in the three wetness regimes, with special
attention to the viscoplastic flow. Here we only examine the
\bead{0.9} beads, since they exhibit viscoplastic continuous flow
in the widest range of parameter space.

%%%%%%%%%%%%%%%%%%%%%%%%%%%%%%%%%%%%%%%%%%%%%%%%%%%%%%%%%%%%%%%%%%%%%%
%%%%                                                              %%%%
\begin{figure}
 \vglue 0.9cm
 \includegraphics[clip,width=\columnwidth]{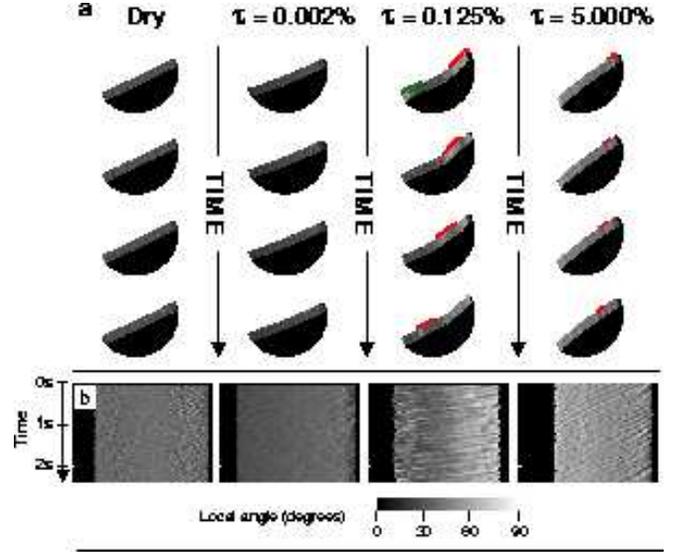}
\caption{\redfig (color) Comparison of the types of continuous flow at \rot{10}
(large beads: \bead{0.9}). \fpc{a} snapshots at time intervals of
$0.067$\,s. The red marks indicate travelling surface features.
\fpc{b} time evolution of the surface angle $\alpha(x,t)$. The
travelling surface features appear as stripes.
 }
 \label{fig_drum_flowtypes}
\end{figure}
%%%%                                                              %%%%
%%%%%%%%%%%%%%%%%%%%%%%%%%%%%%%%%%%%%%%%%%%%%%%%%%%%%%%%%%%%%%%%%%%%%%

\fig{fig_drum_flowtypes} shows the evolution of the surface
profile during different types of continuous flow for \rot{10}. In
the granular regime (e.g. $\tau=0$ and $0.002\,\%$) the surface
profile is practically a straight line with only noise-like
variations. Our resolution is insufficient to resolve the
individually rolling grains. By contrast, in the correlated regime
the surface is clearly concave, and the flow consists of a
succession of clumps moving downhill. From the slope of the
stripes in \ref{fig_drum_flowtypes}\fp{b} we can estimate the
velocity of the moving clumps: $v\approx50-60\,$cm/s. In the
viscoplastic regime the flow is smooth again, and the travelling
surface features demonstrate that there are lasting contacts in
the flowing layer. The slopes of stripes indicate that the flow is
much slower in this regime, $v\approx10-15\,$cm/s. We note that
the rotation rate is the same for all the presented liquid
contents, meaning that the material flux is constant. The
difference in surface velocity thus presumably indicates that the
moving layer is much deeper in case of viscoplastic flow.

%%%%%%%%%%%%%%%%%%%%%%%%%%%%%%%%%%%%%%%%%%%%%%%%%%%%%%%%%%%%%%%%%%%%%%
%%%%                                                              %%%%
\begin{figure}
 \includegraphics[clip,width=6cm]{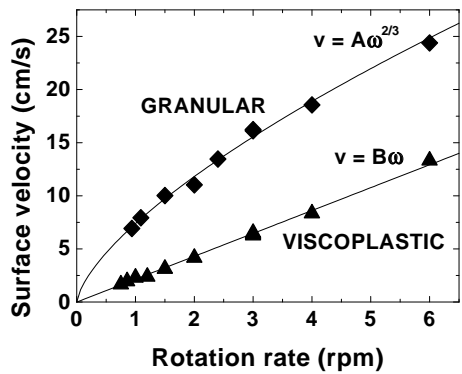}
 \caption{The velocity $v$ at the surface as a function of rotation rate
$\Omega$ for the ordinary granular and the viscoplastic continuous
flow (large beads, \bead{0.9}, the liquid contents are \oil{0.005}
and \oil{0.25} respectively). The continuous lines represent
power-low fits to the data. The linear curve for the viscoplastic
flow indicates that flow depth is independent of the rotation
rate. } \label{drum_flow_vel}
\end{figure}
%%%%                                                              %%%%
%%%%%%%%%%%%%%%%%%%%%%%%%%%%%%%%%%%%%%%%%%%%%%%%%%%%%%%%%%%%%%%%%%%%%%

In order to compare the viscoplastic flow to the ordinary granular
flow, we added some tracer particles to measure the surface
velocity $v$ more accurately as a function of rotation rate,
$\Omega$ (\fig{drum_flow_vel}). The viscoplastic flow is slower by
a factor of two than the granular one, and more importantly, the
two curves correspond to rather different functional forms. The
curvature of the granular curve suggest that at higher rotation
rate the flow depth is increased, as has been previously suggested
\cite{Bonamy01}. Naturally in our finite setup the flow depth
cannot increase infinitely which may explain why our measured
$v(\Omega)$ deviates from $\Omega^{1/2}$ which would be expected
from a linear velocity profile
\cite{Rajchenbach98,Bonamy01,Yamane98}.   On the other hand, the
recent work of Aranson et al. actually predicts a 2/3 power law
\cite{Aranson02,Aranson02a}.
The linear $v(\Omega)$ for viscoplastic flow suggests a {\em constant flow
depth} which is independent of the flow rate. This observation is consistent
with the coherent nature of this type of flow: the flow depth is not determined
by local mechanisms \cite{Rajchenbach02}, but is fixed by the geometry of the
whole system. In order to characterize the viscoplastic flow in greater detail
and check whether the flow depth is really constant we have performed explicit
measurements to determine the extent and shape of the flowing layer. The
results confirmed that the flow depth is almost independent of the rotation
rate {\cite{Tegzes02a}.

\section{Conclusions}
\label{sec_discuss}

We have investigated many features of the dynamic behavior of
wet granular samples in a rotating drum. We found that all these
features are related and in particular all of them reflect the 3
fundamental regimes observed earlier \cite{Tegzes99}.

In the granular regime the cohesive forces are relatively small, and they have
almost no effect in the fluidized flowing layer. Thus only a small increase is
observable in $\tmax$, while $\tr$ hardly changes. The grains are rolling
freely, and avalanche dynamics is dominated by propagating fronts. These form
nearly flat surfaces resulting in relatively small avalanche size fluctuations.
At high rotation rates we observe continuous flow with a fluidized flowing
layer, similarly to surface flows in dry samples.

On the other hand, in the correlated regime the grains stick
together and form clumps. The increased cohesion leads to a large
increase in $\tmax$ and $\tr$. As the free rolling of grains
becomes impossible, the avalanche fronts disappear and steep,
rugged, and usually concave surfaces are formed. Variations of the
surface shape lead to wide avalanche size distributions. In this
regime the continuous flow consists of a succession of falling
clumps, thus the hysteresis between avalanching and continuous
flow disappears.

Finally, in the viscoplastic regime the flow becomes coherent over
the entire sample, and $\tmax$ and $\Dt$  decrease slightly due to
lubrication and viscous effects. The reduced fluctuations lead to
narrower avalanche size distributions and the formation of robust
patterns that are reproduced at the end of each avalanche. The
transition to continuous flow happens at relatively low rotation
rates, and the hysteresis reappears. The coherence of the
viscoplastic continuous flow is manifested in traveling surface
features and a velocity independent flow depth.

While the flows we observe appear to have analogies with geological events, it
is important to note that real geological materials usually consist of
polydisperse irregular particles often with very high ($\tau\approx100\%$)
water contents and that the scaling of our system to geological lengthscales is
non-trivial (indeed, based our on earlier work we expect that all of the results will be affected by the size of the container \cite{Tegzes99}).  Furthermore, avalanche studies in real soil have demonstrated
additional phenomena associated with soil saturation \cite{Iverson00}.
Interestingly we still recover some of the basic dynamical processes in our
model system, which should aid the description of
qualitatively different flow behaviors in the framework of a single model
\cite{Nase01,Gray01}.

The changes in the dynamic behavior with wetting are associated with the
increasingly coherent nature of the flow, i.e. the formation of coherently
moving clusters -- clumps -- due to the increased cohesion and viscous effects.
Within a cluster, local velocity fluctuations should be suppressed, and thus
the local granular temperature ($T = \avg{ v^2 } - \avg{ v }^2$) should
approach zero, but the clusters themselves both form and break apart during an
avalanche process in a finite container. An important theoretical question
raised by our data is how a length scale describing the size of the clumps may
emerge from a granular flow model, and how such a length scale would vary with
the type of media, the total size of the granular sample, the nature of intergranular adhesion, viscosity of the
fluid, and the type of granular flow. While the present experiments have raised
these questions, further experimental studies expanding the investigated phase
space as well as computational modelling will almost
certainly be necessary before a complete understanding is approachable.

\begin{acknowledgments}

We gratefully acknowledge helpful discussions with J. Banavar, A.-L.
Barab\'asi, and Y. K. Tsui. We are also grateful for support from the Petroleum
Research Fund and NASA Grant NAG3-2384. P. T. and T. V. are grateful for the
partial support from OTKA Grant No. T033104.

\end{acknowledgments}

%\bibliography{granu}% Produces the bibliography via BibTeX.

\end{document}